\documentclass[12pt,preprint]{aastex}

\begin{document}

\received{}
\revised{}
\accepted{}

\shortauthors{Temim et al.}

\shorttitle{The ``Crab Nebula with Spitzer''}

\title{SPITZER SPACE TELESCOPE INFRARED IMAGING AND SPECTROSCOPY OF
THE CRAB NEBULA}

\author{TEA TEMIM\altaffilmark{1}, ROBERT D. GEHRZ\altaffilmark{1}, CHARLES E.
WOODWARD\altaffilmark{1}, THOMAS L. ROELLIG\altaffilmark{2}, NATHAN
SMITH\altaffilmark{3,6},  LAWRENCE R. RUDNICK\altaffilmark{1},ELISHA  F.
POLOMSKI\altaffilmark{1},  KRIS  DAVIDSON\altaffilmark{1}, LUNMING
YUEN\altaffilmark{4}, TAKASHI ONAKA\altaffilmark{5}}

\altaffiltext{1}{Department of Astronomy, School of Physics and Astronomy, 116  Church Street,
S.E., University of Minnesota, Minneapolis, MN 55455,  \it{gehrz@astro.umn.edu},
\it{ttemim@astro.umn.edu}, \it{chelsea@astro.umn.edu},  \it{elwood@astro.umn.edu},
\it{larry@astro.umn.edu},  \it{kd@astro.umn.edu}}

\altaffiltext{2}{NASA Ames Research Center, MS 245-6, Moffett Field, CA 94035-1000,
\it{Thomas.L.Roellig@nasa.gov}}

\altaffiltext{3}{Center for Astrophysics and Space Astronomy, University of Colorado, 389
UCB, Boulder, CO 80309, \it{nathans@casa.colorado.edu}}

\altaffiltext{4}{Technosciences Corp., MS 245-6,  Moffett Field, CA 94035,
\it{lyuen@mail.arc.nasa.gov}}

\altaffiltext{5}{University of Tokyo, 7-3-1 Hongo, Bunkyo-ku, Tokyo 113-0033  Japan,
\it{onaka@astron.s.u-tokyo.ac.jp}}

\altaffiltext{6}{Visiting Astronomer, Cerro Tololo Inter-American
Observatory, National Optical Astronomy Observatory (NOAO), operated
by the Association of Universities for Research in Astronomy (AURA),
Inc., under cooperative agreement with the National Science
Foundation (NSF).}

\begin{abstract}
We present 3.6, 4.5, 5.8, 8.0, 24, and 70 $\micron$ images of the
Crab Nebula obtained with the Spitzer Space Telescope IRAC and MIPS
cameras, Low- and High-resolution Spitzer IRS spectra of selected positions
within the nebula, and a near-infrared ground-based image made in
the light of [Fe II]1.644 $\micron$. The 8.0 $\micron$ image, made
with a bandpass that includes [Ar II]7.0 $\micron$, resembles the
general morphology of visible H$\alpha$ and near-IR [Fe II] line
emission, while the 3.6 and 4.5 $\micron$ images are dominated by
continuum synchrotron emission. The 24 $\micron$ and 70 $\micron$
images show enhanced emission that may be due to line emission or
the presence of a small amount of warm dust in the nebula on the
order of less than 1\% of a solar mass. The ratio of the 3.6 and 4.5
$\micron$ images reveals a spatial variation in the synchrotron
power law index ranging from approximately 0.3 to 0.8 across the
nebula. Combining this information with optical and X-ray
synchrotron images, we derive a broadband spectrum that reflects the
superposition of the flatter spectrum jet and torus with the steeper
diffuse nebula, and suggestions of the expected pileup of
relativistic electrons just before the exponential cutoff in the
X-ray. The pulsar, and the associated equatorial toroid and polar
jet structures seen in Chandra and HST images \citep{Hes02} can be
identified in all of the IRAC images. We present the IR photometry
of the pulsar. The forbidden lines identified in the high resolution
IR spectra are all double due to Doppler shifts from the front and
back of the expanding nebula and give an expansion velocity of $\approx 1264$
km s$^{-1}$.

\end{abstract}


\section{INTRODUCTION \label{intro} }

In their seminal paper ``Synthesis of the Elements in Stars,'' E. M.
Burbidge, G. R. Burbidge, W. A. Fowler, and F. Hoyle (1957)
\nocite{Bur57} described how primordial hydrogen is converted into
the other elements by nucleosynthesis in stellar interiors and
stellar explosions. Massive stars play a particularly important
role in the production of nuclei up to the iron peak during main
sequence and post main sequence nucleosynthesis. The Type II
Supernova (SN) explosions that result from the collapse of the their
iron cores produce the rest of the heavy elements by rapid neutron
capture in the expanding ejecta. Thus, SN ejecta enrich the
chemical content of the interstellar clouds from which new stars
continually form, and it is believed that the contents of the
pre-solar nebula were profoundly affected by one or more such events
\citep[see][]{Cla82}.

In particular, the ability of SN explosions to eject copious
quantities of carbon and other condensable metals has long led to
speculation that SNs are capable of producing vast quantities of
stardust as their ejecta cool \citep{Cla82, Geh87, Dwe88, Geh88}.
Clayton has proposed that such grains may survive today in meteorite
inclusions. The infrared (IR) is an ideal spectral region in which
to test these theories, because dust grains emit strongly in the IR
as do forbidden emission lines from condensable metals that remain
in the gas phase. On the other hand, previous IR studies that have
failed to reveal evidence for any large amount of dust in SN
remnants (SNRs) have provided us with somewhat of a mystery
\citep{Geh90, Are99, Dou01, Dwe04, Gre04}.

We have used Guaranteed Observation Time (GTO) on NASA's new Spitzer
Space Telescope \citep{Wer04} to obtain IR images and spectroscopy
of the Crab Nebula, formed by a supernova explosion in 1054 AD. It
is one of the youngest known SNRs and is one of the most studied
objects in the Galaxy. The SN ejecta in the Crab are concentrated
in filaments of ionized gas that produce emission line spectra
\citep[see][]{Dav85}. One long-standing puzzle regarding the Crab
is the lack of a visible blast wave expected from a massive star
explosion \citep{Sew06}. Previous IR observations have indicated
that there is a paucity of dust in the Crab. Infrared Astronomical
Satellite (IRAS) observations revealed an IR excess at
wavelengths longer than 12 $\micron$ that was attributed to thermal
emission from 0.005-0.03 M$_{\odot}$ of dust \citep{Mar84}. Further
evidence for dust in the Crab Nebula was found in the form of
optical extinction in the filaments \citep{Hes90, Fes90, Bla97,
San98}. In more recent studies using the Infrared Space Observatory
(ISO), \citet{Dou01} report no evidence of spectral features from
dust emission in the ISOCAM spectra, while \citet{Gre04} calculate
an upper limit of 0.02 M$_{\odot}$ of warm dust from the far-IR
excess seen by ISOPHOT.

The Crab Nebula is one of the brightest synchrotron sources in the
Galaxy. It is a prototype of a class of SNRs called the
filled-center SNRs or ``plerions,'' that are powered by a central
pulsar. A large fraction of the Crab pulsar's spin down luminosity
is converted into the nebular synchrotron luminosity, from radio
through gamma-rays. The details of this conversion are quite
uncertain. The \citet{Ken84} steady-state spherical MHD model still
provides the best description of the optical and X-ray profiles, but
cannot tie them together with the radio synchrotron emission.

The unprecedented sensitivity of the IR imagers and spectrometers of
the Spitzer Space Telescope present us with an unparalleled
opportunity to search for dust and forbidden line emission in the
Crab Nebula and study in detail the spatial variations of
synchrotron emission across the remnant. In this paper, we present
3.6, 4.5, 5.8, 8.0, 24, and 70 $\micron$ images of the Crab Nebula
obtained with the Spitzer Space Telescope IRAC and MIPS cameras, IRS
spectra of selected positions within the nebula, and a near-IR
ground-based image made in the light of [Fe II]1.644 $\micron$.

\section{OBSERVATIONS AND DATA REDUCTION\label{obsv}}

Observations of the Crab Nebula were made using the Infrared Array
Camera (IRAC) aboard the Spitzer Space Telescope \citep{Wer04} at
3.6, 4.5, 5.8 and 8.0 $\micron$ and the Multiband Imaging Photometer
for Spitzer (MIPS) at 24 and 70 $\micron$ as part of the Gehrz
Guaranteed Time Observing Program (GGTOP, Program ID: 130). High and
low resolution Spitzer 5-40 $\micron$ spectra of selected regions of
the Crab Nebula were made using the Infrared Spectrometer (IRS)
under the Roellig Guaranteed Time Observing Program (RGTOP, Program
ID: 24).

\subsection{IRAC IMAGES \label{obsv_IRAC}}

The IRAC \citep{Faz04} observations were made on 2004 March 6, under the
AORKey 6588928,  using 12 second exposures and the High Dynamic Range
(HDR) mode at 4 dither positions obtained in a cycling dither pattern.
The Basic Calibrated Data (BCD) products from the Spitzer Science Center
(SSC) pipeline version S11.0.2, calibrated in units of MJy per steradian,
were used in our post pipeline processing.

Post BCD processing was conducted using the 101504 version of the
SSC Legacy MOPEX software. The reduction with MOPEX consisted of
three steps: Cosmetic Fix, Background Matching, and Mosaicker. The
field of view of the final mosaics is approximately 6.0\arcmin $\times$ 6.0\arcmin
with a final pixel scale of 0.86$\arcsec$ per pixel.

\subsection{MIPS IMAGES \label{obsv_MIPS}}

The MIPS \citep{Rie04}  24 $\micron$ and 70 $\micron$ observations
of the Crab Nebula were carried out on 2004 March 14, under the AORKey 6588672,
 using an exposure time of 3 seconds and 10 seconds, respectively. The data
was processed with the SSC pipeline version S11.4.0. The 24
$\micron$ BCD data was corrected for image distortions caused by the
focal plane and mosaicked with Image Reduction and Analysis Facility
(IRAF)\footnote{IRAF is distributed by the National Optical
Astronomy Observatories, which are operated by the Association of
Universities for Research in Astronomy, Inc., under cooperative
agreement with the National Science Foundation}. The 70 $\micron$
images were mosaicked with MOPEX and have a pixel scale of
4.0$\arcsec$ per pixel. The pixel scale of the final MIPS 24
$\micron$ mosaic is 2.5$\arcsec$ per pixel. The summary of the IRAC
and MIPS data is listed in Table \ref{irac_mips}.

\subsection{IRS SPECTRA \label{obsv_IRS}}

\begin{figure}
\epsscale{1.0} \plotone{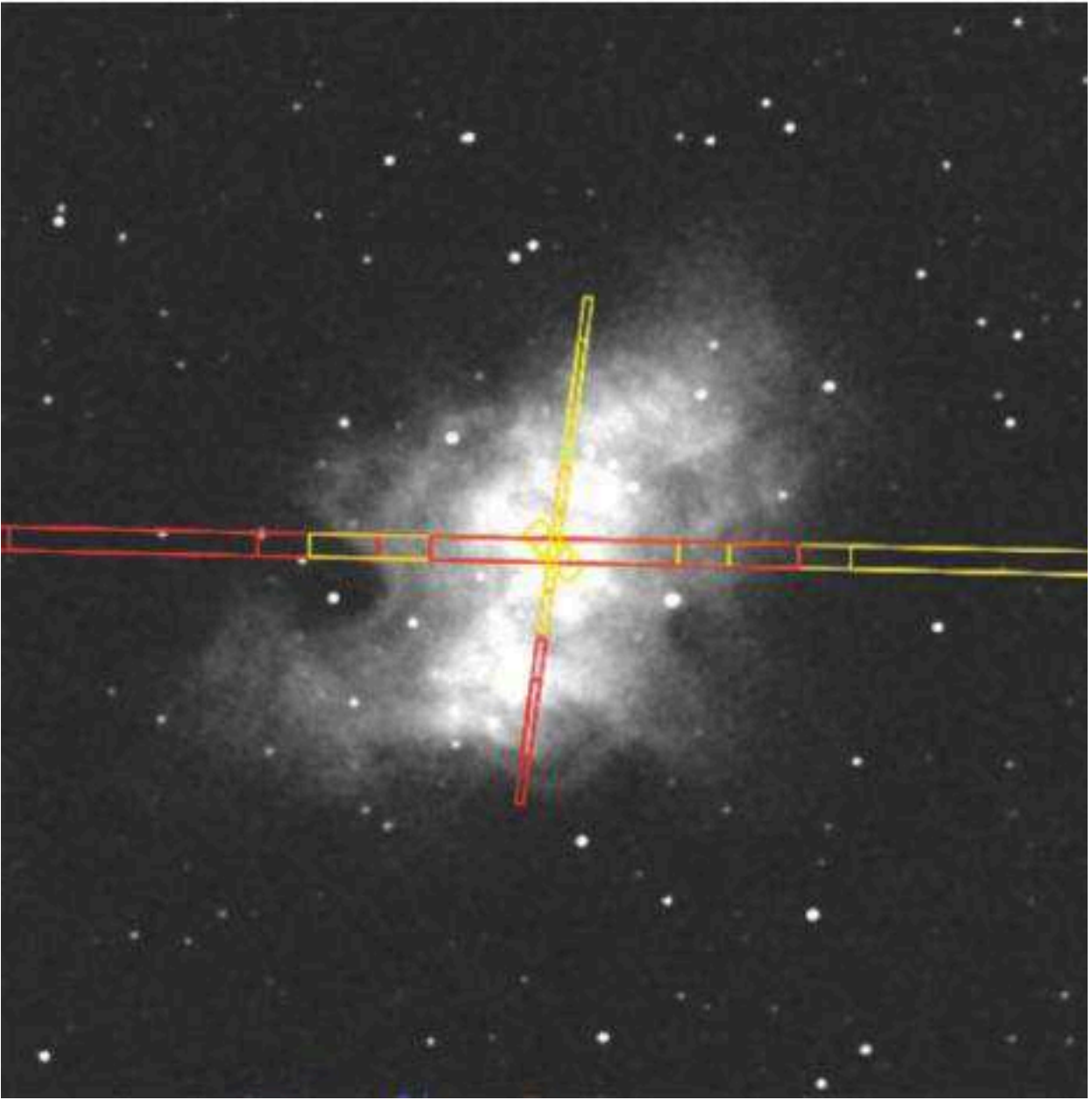} \caption{\label{TLR-1} A
2MASS K-band image of the Crab nebula, with the IRS slits overlaid
at the time of the observations reported here. This figure was
produced by the SPOT software at the Spitzer Science Center and
shows each IRS module's aperture orientation in the two nod
positions along the slits. The long and wide aperture running
horizontally belongs to the Long-Low module, while the much thinner
aperture running more vertically belongs to the Short-Low aperture.
The high-resolution modules' apertures are much shorted and only the
Long-High aperture is readily visible in this figure.}
\end{figure}

Observations of the Crab nebula were obtained with all four modules
of the IRS instrument \citep{Hou04} on 2004 March 5, under the
AORKey 3857664. The IRS observations were taken in staring mode
centered at the position RA 5h34m31.02s, Dec 22d01\arcmin10.1\arcsec (J2000).
This position, designated as IRS-Tgt-Cntr in the discussions below,
is located in the center of the nebula, near but not coincident with
the pulsar position located at RA 5h34m31.97s, Dec 22d00\arcmin52.1\arcsec
(J2000). Since the four IRS modules all have different slit sizes
and orientations, this position in the nebula was the only one in
common to all the modules. For the two low-resolution IRS modules
the long slits allowed sampling of other regions of the nebula, as
shown in Figure \ref{TLR-1}. Integration times, wavelength coverage,
and the spectral resolution for each IRS module is given in Table
\ref{irs}. We also report data from other locations in the nebula
sampled by the long slits in the low-resolution modules. Since the
IRS staring mode involves nodding between two positions in each
slit, some of these regions have the same integration times as those
reported above, while others have half the time. Further explanation
of how nodding along the slits is implemented in the IRS staring
mode can be found in the relevant section of the Spitzer Observers
Manual\footnote{See http://ssc.spitzer.caltech.edu/documents/SOM /
for the most current version of the Spitzer Observers Manual}.

\begin{figure}
\epsscale{1.0} \plotone{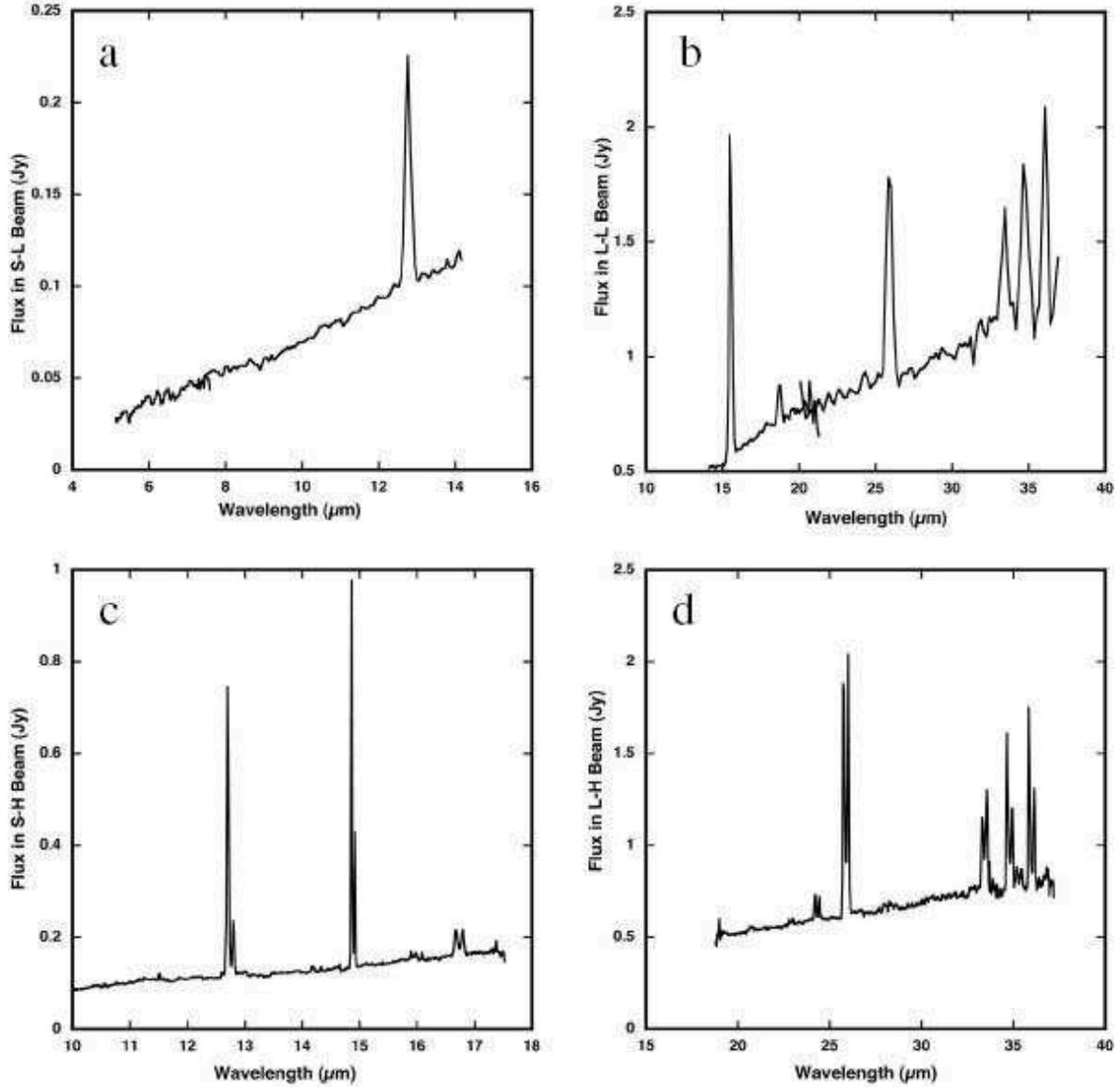} \caption{\label{TLR-2}
Spitzer IRS spectra of the Crab nebula, taken at the IRS-Tgt-Cntr
position. Panel (a): Short-Low module with a 3.6\arcsec x 9.0\arcsec extracted
beam size; (b) Long-Low module with a 10.5\arcsec x 25.5\arcsec extracted beam
size; (c) Short-High module with a 4.7\arcsec x 11.3\arcsec extracted beam size,
and (d): Long-High module with a 11.1\arcsec x 22.3\arcsec extracted beam size.
Identifications of the numerous emission lines are given in Tables
\ref{sh} and \ref{lh}.}
\end{figure}

The IRS data were reduced using the SMART data reduction package
described in \citet{Hig04}. Due to the large spatial extent of the
Crab nebula with respect to the IRS slit dimensions there was no
blank sky available for zodiacal light subtraction. In this region
of the sky the zodiacal sky background levels are roughly one-fourth
of the combined emission from the Crab Nebula and zodiacal light at
the worst-case wavelengths around 15 $\micron$. As a result, the
data reported here have a small contribution from the Zodiacal
background that must be accounted for. The observed spectra from
each of the IRS modules for the IRS-Tgt-Cntr position are shown in
Figure \ref{TLR-2}. In these spectra the high-resolution module
extractions were both full-aperture, while for the low-resolution
module's long slits the extractions were performed in 5 pixel wide
sub-slits centered at the location of IRS-Tgt-Cntr. The effects of
the different slit widths and extraction aperture sizes are
immediately noticeable in the relative strengths of the emission
baselines in the spectra in Figure \ref{TLR-2}.

\subsection{NEAR-IR IMAGING \label{obs_nearIR}}

Since [Fe~{\sc ii}] 25.99 $\micron$ emission may contribute to the
emission morphology at 24~$\micron$ (see below), it will be useful
to compare MIPS to a ground-based image of the [Fe~{\sc ii}]
1.644~$\micron$ emission line. We obtained images of the Crab Nebula
on 2001 March 9 using the Ohio State IR Imaging Spectrometer
(OSIRIS)\footnote{OSIRIS is a collaborative project between the Ohio
State University and Cerro Tololo Inter-American Observatory (CTIO)
and was developed through NSF grants AST 90-16112 and AST 92-18449.
CTIO is part of the National Optical Astronomy Observatory (NOAO),
based in La Serena, Chile. NOAO is operated by the Association of
Universities for Research in Astronomy (AURA), Inc. under
cooperative agreement with the National Science Foundation} mounted
on the NOAO Cerro Tololo Inter-American Observatory (CTIO) 1.5 m
telescope (see panel (a) of Figure \ref{Fe2comp}). OSIRIS has a
1024$\times$1024 NICMOS3 array, with a pixel scale of 1$\farcs$153
arcsec using the f/2.8 camera. Only a portion of the array is
illuminated, yielding a field of view of 11$\arcmin$. We used a
narrow ($\Delta\lambda$/$\lambda\simeq$1\%) filter to image the
extended [Fe~{\sc ii}] 1.644~$\micron$ line emission from the Crab.
Six individual exposures of 120~s each were sky subtracted using a
median of similar images centered at a position 10$\arcmin$ south of
the Crab (with slight positional offsets between each), and then
shifted and co-added.

We used H-band fluxes from the 2MASS point source catalog for
several stars in the field in order to flux calibrate the resulting
[Fe~{\sc ii}] image. We also used the H-band image from the 2MASS
survey to subtract the continuum emission from the synchrotron
nebula that is included in the [Fe~{\sc ii}] filter, in order to
produce a true image of the line emission.  While the H-band image
is not a pure continuum image (the [Fe~{\sc ii}] 1.644 $\micron$
line is included in the H band), the broadband image is dominated by
synchrotron continuum and the thermal filaments are barely seen;
based on the flux of the [Fe~{\sc ii}] line, we estimate that it
contributes only 2--3\% of the total flux in the broad H-band
filter.

\section{ANALYSIS AND RESULTS \label{disc}}

\subsection{MULTI-WAVELENGTH MORPHOLOGY \label{multi}}

\subsubsection{IRAC and MIPS morphology \label{morph}}

\begin{figure}
\epsscale{0.6} \plotone{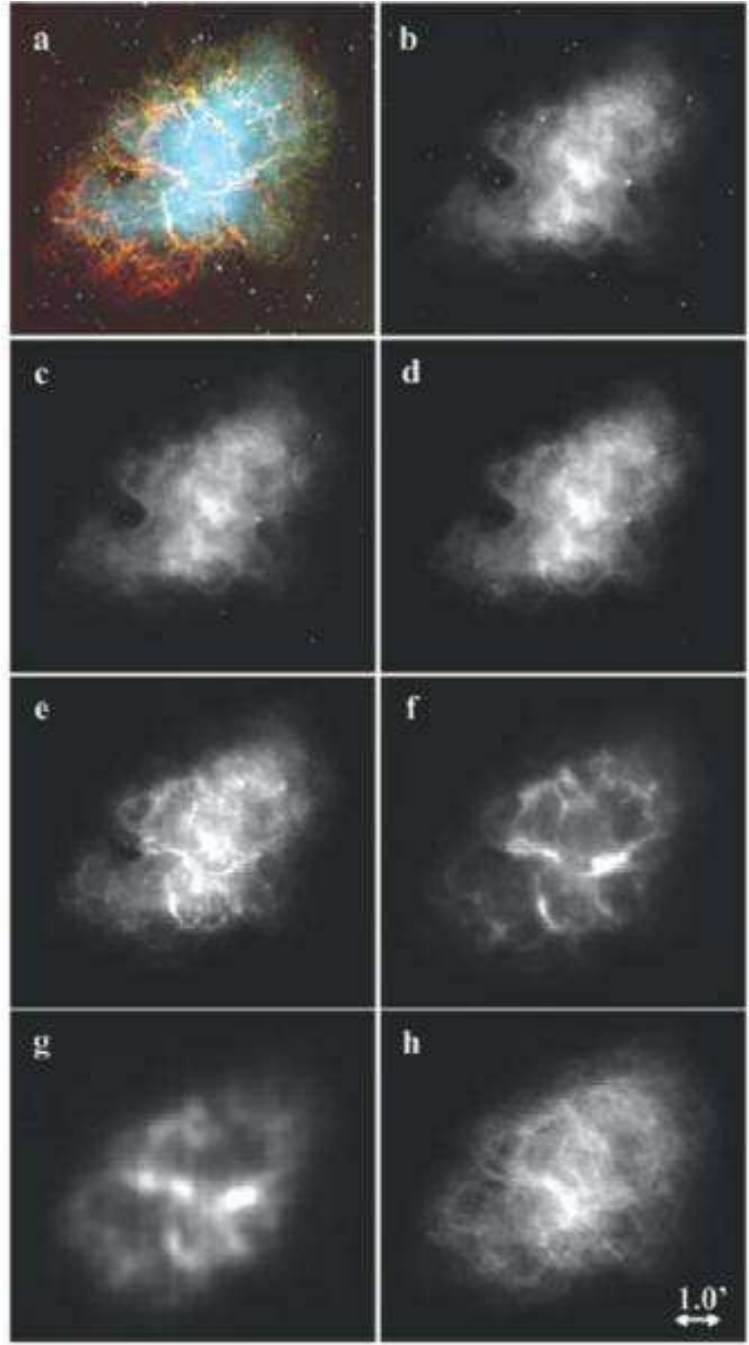} \caption{\label{nebulapanel} A
comparison of Spitzer IR images of the Crab Nebula with visual and
radio images showing that the smooth synchrotron component and the
filamentary component dominate the emission at different
wavelengths.  Panel (a) is a 3 color visual press release image
reproduced courtesy of the European Southern Observatory. Panels
(b), (c), and (d) show IRAC 3.6 $\micron$, 4.5 $\micron$, and 5.8
$\micron$ images that trace out the synchrotron component. Panels
(e), (f), and (g) show IRAC 8.0 $\micron$ and MIPS 24 $\micron$ and
70 $\micron$ images that show filamentary structures dominated by
strong forbidden line emission. Panel (h) shows that the 5 GHz radio
image traces out both the smooth synchrotron component and thermal
bremsstrahlung in the filaments \citep[image reproduced courtesy of
NRAO/AUI and][]{Bie01}}
\end{figure}

\begin{figure}
\epsscale{1.0} \plotone{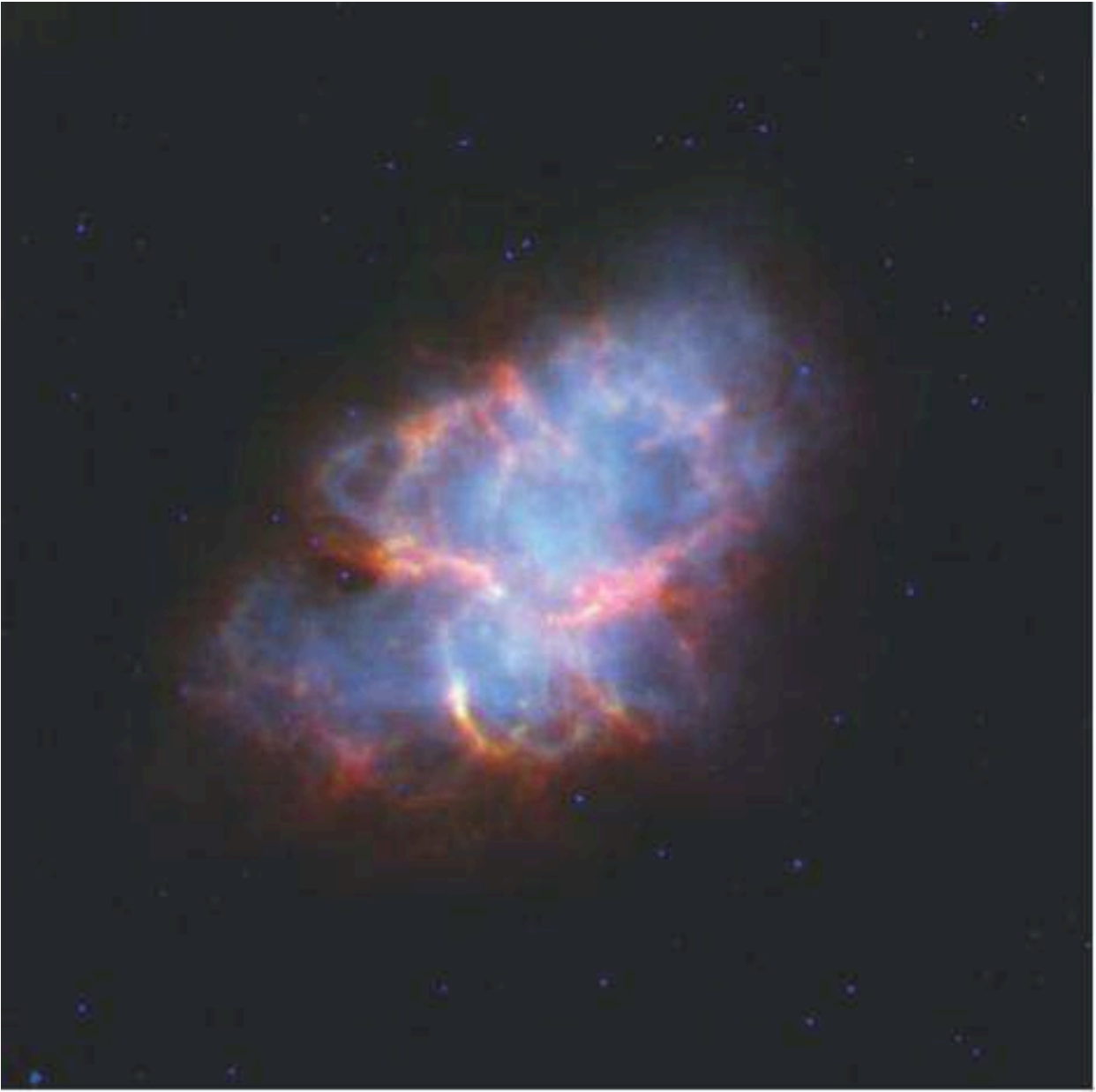} \caption{\label{3color} Spitzer
IRAC/MIPS three-color composite image.  Blue (3.6 $\micron$)
emission maps out the synchrotron component, green (8.0 $\micron$)
maps out [Ar II] 7.0 $\micron$  emission, and red (24 $\micron$)
maps out [O IV] 25.9 $\micron$ emission. The color tables are linearly
proportional to intensity.}
\end{figure}

IRAC and MIPS images of the Crab Nebula are displayed in Figure
\ref{nebulapanel}. Panel (a) is a three color visual press release
image reproduced courtesy of the European Southern Observatory,
showing the H$\alpha$ filaments that trace out regions of strong
hydrogen recombination line emission. The smooth blue background is
due to synchrotron radiation from relativistic electrons ejected by
the neutron star central engine. Panels (b), (c), and (d) of Figure
\ref{nebulapanel} show IRAC 3.6 $\micron$, 4.5 $\micron$, and 5.8
$\micron$ images whose emission is dominated by synchrotron
emission. Panel (e) of Figure \ref{nebulapanel} shows the IRAC 8.0
$\micron$ image, and panels (f) and (g) are MIPS 24 $\micron$ and 70
$\micron$ images. Images in panels (e), (f), and (g) all show
filamentary structure that is dominated by strong forbidden line
emission and correlates very strongly with the recombination line
filaments traced out in the visual H$\alpha$ image of panel (a). The
NRAO 5 GHz radio image \citep{Bie01} is displayed in panel (h) for comparison and
it shows both the smooth synchrotron component and the thermal
bremsstrahlung in the filaments. The IRAC and MIPS three-color
composite image is displayed in Figure \ref{3color}. Blue represents
the IRAC 3.6 $\micron$ morphology that traces out the synchrotron
emission, while the green (8.0 $\micron$) and red (24 $\micron$) map
out the [Ar~{\sc ii}] 7.0 $\micron$ and [O~{\sc iv}] 26 $\micron$
(see Section 3.3) emission respectively and trace the filamentary
structure seen in H$\alpha$. While the south
filament is bright in both green and red, the equatorial filament
is dominated by MIPS 24 $\micron$ emission which may suggest the
presence of large dust grains (see Section 4.2).

\begin{figure}
\epsscale{1.0} \plotone{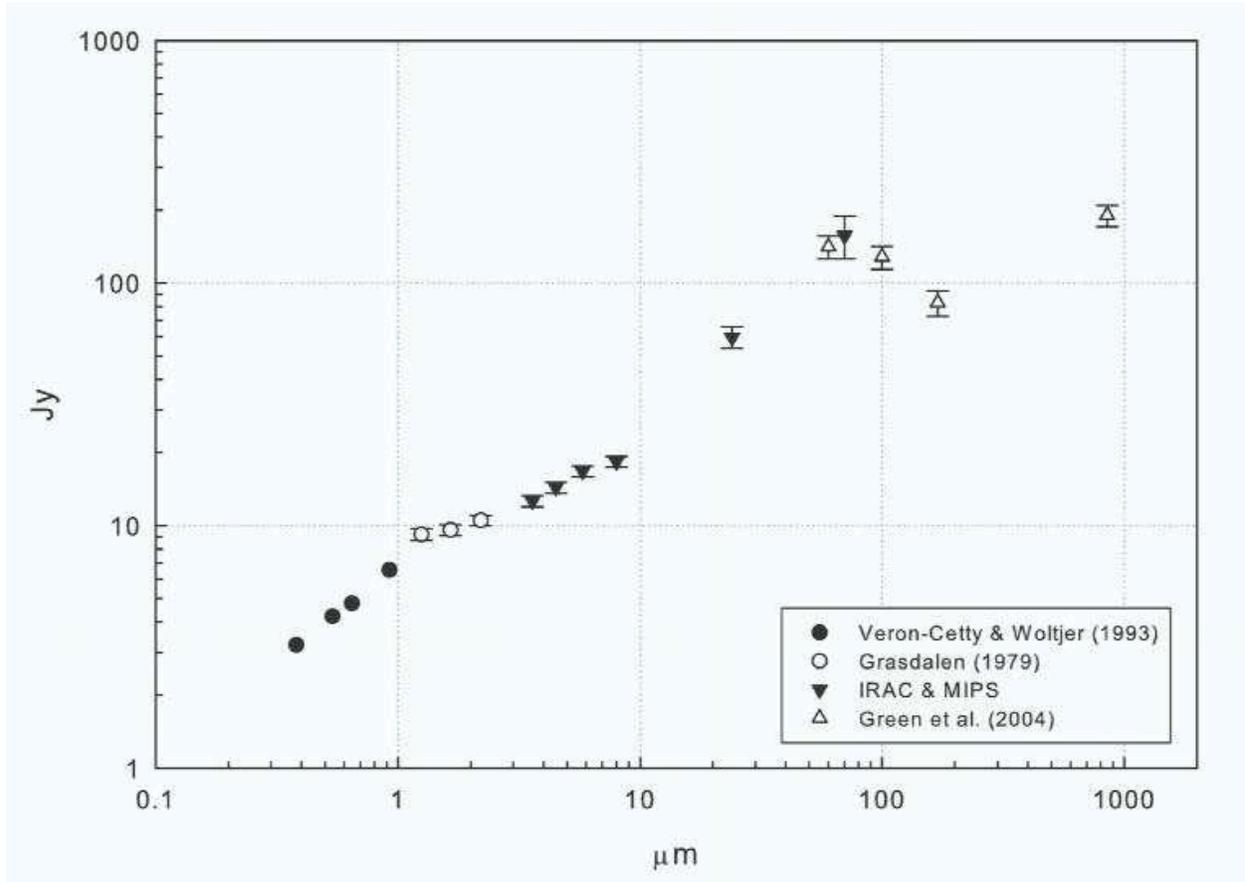} \caption{\label{nebulaflux}
Integrated flux density of the Crab Nebula. IRAC wavelengths are
corrected for extinction using A$_V$ of 1.5 and the extinction
values from \cite{Rie85}. Extended emission correction was also
applied to the IRAC fluxes \citep{Rea05}. IRAC and MIPS errors are
dominated by flux calibrations uncertainties. The error bars
include calibration uncertainties of 5\% for IRAC wavelengths, 10\% for
MIPS 24 $\micron$, and 20\% for MIPS 70 $\micron$.}
\end{figure}

The IRAC and MIPS integrated flux densities of the Crab Nebula are
displayed in Figure \ref{nebulaflux} and listed in Table
\ref{irac_mips}. Uncertainties in the table do not reflect IRAC and
MIPS calibration uncertainties. Extended emission correction was
applied to the IRAC integrated flux densities of the nebula based on
Reach et al. (2005). Figure \ref{nebulaflux} also includes fluxes in
the visible and near IR wavelengths for comparison \citep{Ver93,
Gra79, Gre04}. It can be seen from the plot that the IRAC and MIPS
fluxes agree well with previous data. The IRAC points trace the
synchrotron continuum with a spectral index of 0.5, as derived by
\citet{Dou01}. The excess radiation seen at 24-100 $\micron$ above
the synchrotron continuum may indicate the presence of a small
amount of warm dust in the form of relatively large grains. The
excess at 24 $\micron$ may also partly be due to forbidden line
emission from [O~{\sc iv}]. Some of the excess in the MIPS 70
$\micron$ filter may be due to [O~{\sc i}] 63 $\micron$, [O~{\sc
iii}] 52 $\micron$ and [O~{\sc iii}] 88 $\micron$ \citep{Gre04}.

\subsubsection{Central Nebula and the Crab Pulsar \label{puls}}

\begin{figure}
\epsscale{0.8} \plotone{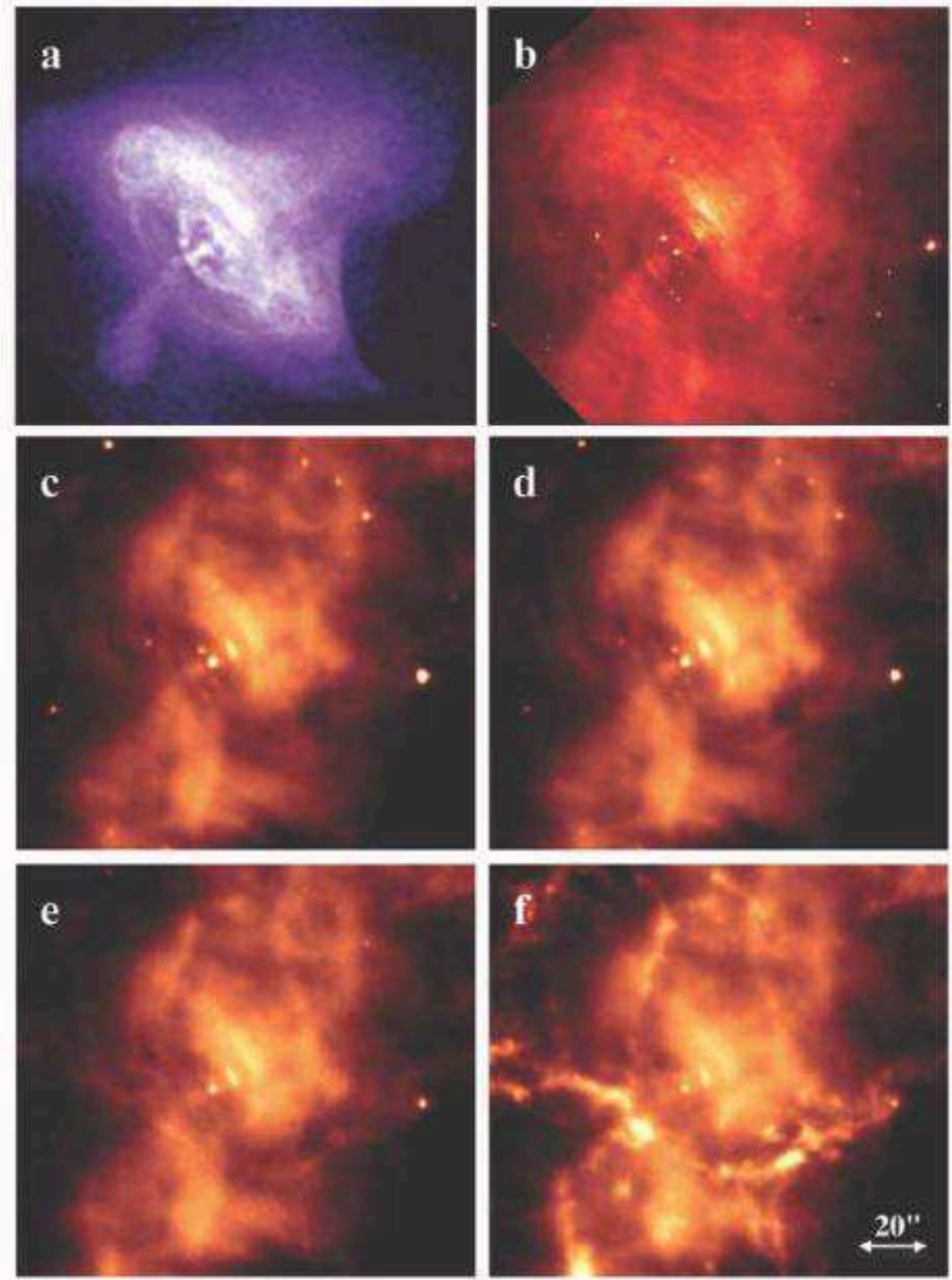} \caption{\label{pulsarpanel} Detail
of the pulsar central engine and its equatorial torus and polar jets
as seen at six wavelengths. The CHANDRA X-ray ACIS-S and
HST optical images are reproduced in panels (a) and (b)
\citep{Hes02}.  Panels (c), (d), (e), and (f) show the 3.6
$\micron$, 4.5 $\micron$ , 5.8 $\micron$, and 8.0 $\micron$ IRAC
images.}
\end{figure}

Close up IRAC images of the central part of the Crab Nebula are
shown in Figure \ref{pulsarpanel}. Panel (a) is the CHANDRA X-ray
Observatory ACIS-S image \citep{Hes02}, panel (b) is the HST
optical image \citep{Hes95}, and panels (c) through (f) are IRAC
images in order of increasing wavelength. The equatorial torus and
polar jets of the X-ray image are clearly visible at all IRAC
wavelengths. The position of the point source in the center of IRAC
images is coincident with the position of the Crab pulsar in the HST
image. The pulsar is not detected in the MIPS images. Additional
features identified in the HST image by \citet{Hes95} are also
visible at IRAC wavelengths, including the knot located 3\farcs8
southeast of the pulsar, aligned with the jet, and the bright arcs
(wisps) located 7\farcs3 northwest of the pulsar.

\begin{figure}
\epsscale{1.0} \plotone{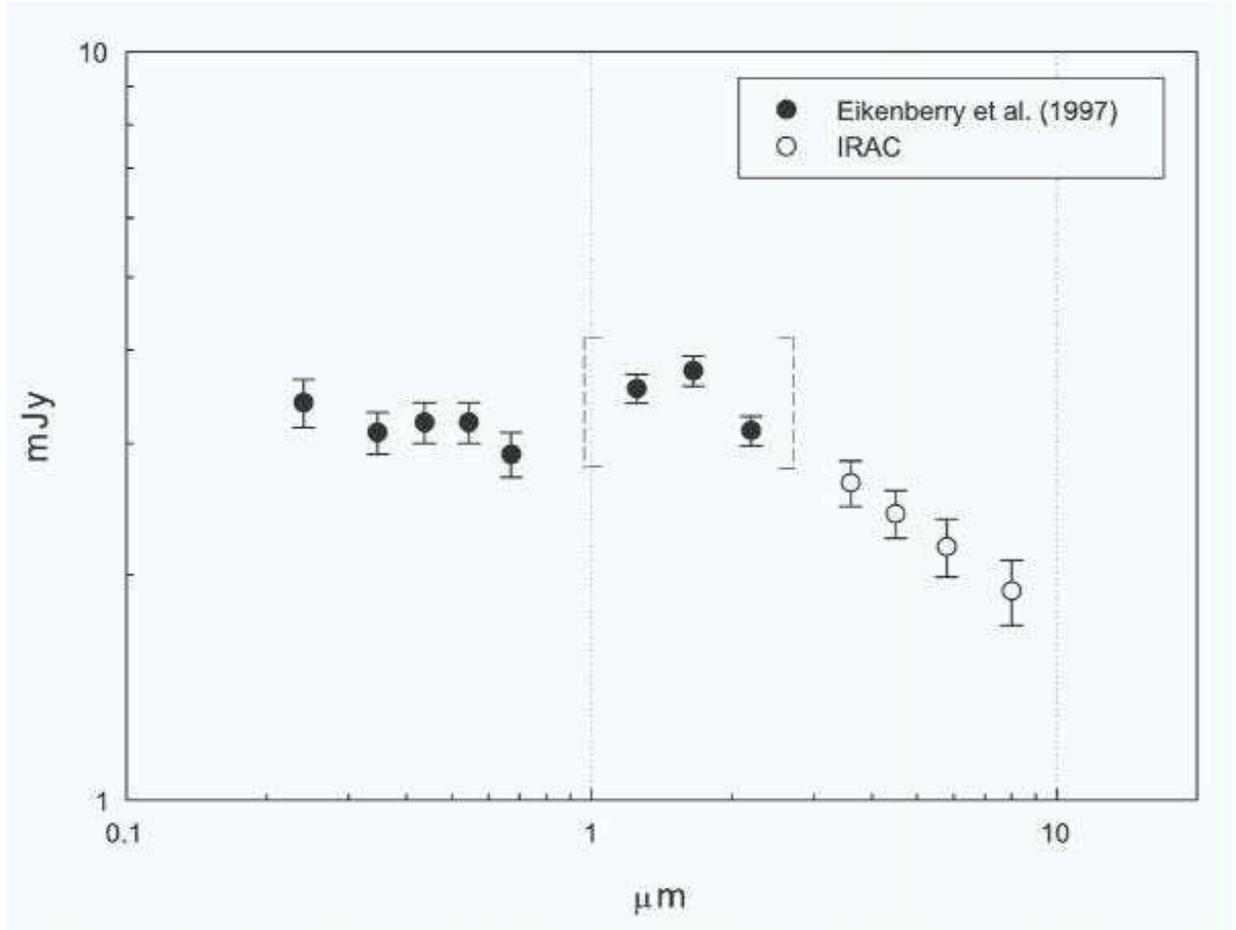} \caption{\label{pulsarflux} The
extinction corrected spectral energy distribution of the Crab
Pulsar. Extinction correction was computed using \citet{Rie85} and
an A$_V$ value of 1.5. The fluxes have been calculated using an
aperture size of 3 pixels and a sky background annulus from 3-7
pixels. Appropriate aperture correction was also applied. The IRAC
error bars include calibration uncertainties of roughly 5\%, but
flux values in this plot likely have additional uncertainties due to
nebular emission around the pulsar. The near IR and visible data are
from Table 5 of \citet{Eik97}, where the visible data come from
\citet{Per93}. The brackets in the plot indicate that there is
significant scatter in the near IR measurements in the literature
\citep[see][]{OCon05}}
\end{figure}

The spectral energy distribution (SED) of the pulsar is shown in
Figure \ref{pulsarflux}. The aperture photometry was performed on
Spitzer images using an aperture radius of three pixels, a
background anulus of 3-7 pixels, and median sky subtraction. The
appropriate aperture correction was applied to the resulting fluxes
(see The Spitzer Observers Manual). The fluxes likely have
additional uncertainties due to high nebular emission around the
pulsar, including the IR knot located 0\farcs6 from the pulsar
identified by \citet{Hes95} and discussed by \citet{Sol03}. Visible
and near IR points from Table 5 of \citet{Eik97}, in which the
visible data comes from \citet{Per93}, are included in Figure
\ref{pulsarflux} for comparison. While the spectrum of the pulsar in
the UV and optical bands remains flat, the IRAC data show a falloff
in the IR with a spectral index of -0.4 $\pm$ 0.1. \citet{Sol03}
found a spectral slope of -0.31 $\pm$ 0.02 in the near-IR,
consistent with the expected index of -1/3 in the low frequency
region of the pulsar spectrum. \citet{OCon05} discuss a possible
rollover due to synchrotron self-absorption in the near IR spectrum
of the Crab pulsar, but it is still unclear if this is the cause of
the falloff seen in in the IRAC data. Table \ref{irac_mips}
summarized the pulsar measurements and does not include calibration
uncertainties of roughly 5\% at each IRAC wavelength.

\subsubsection{IRS line emission/IRAC image morphology comparisons \label{linecomp}}

\begin{figure}
\epsscale{0.9} \plotone{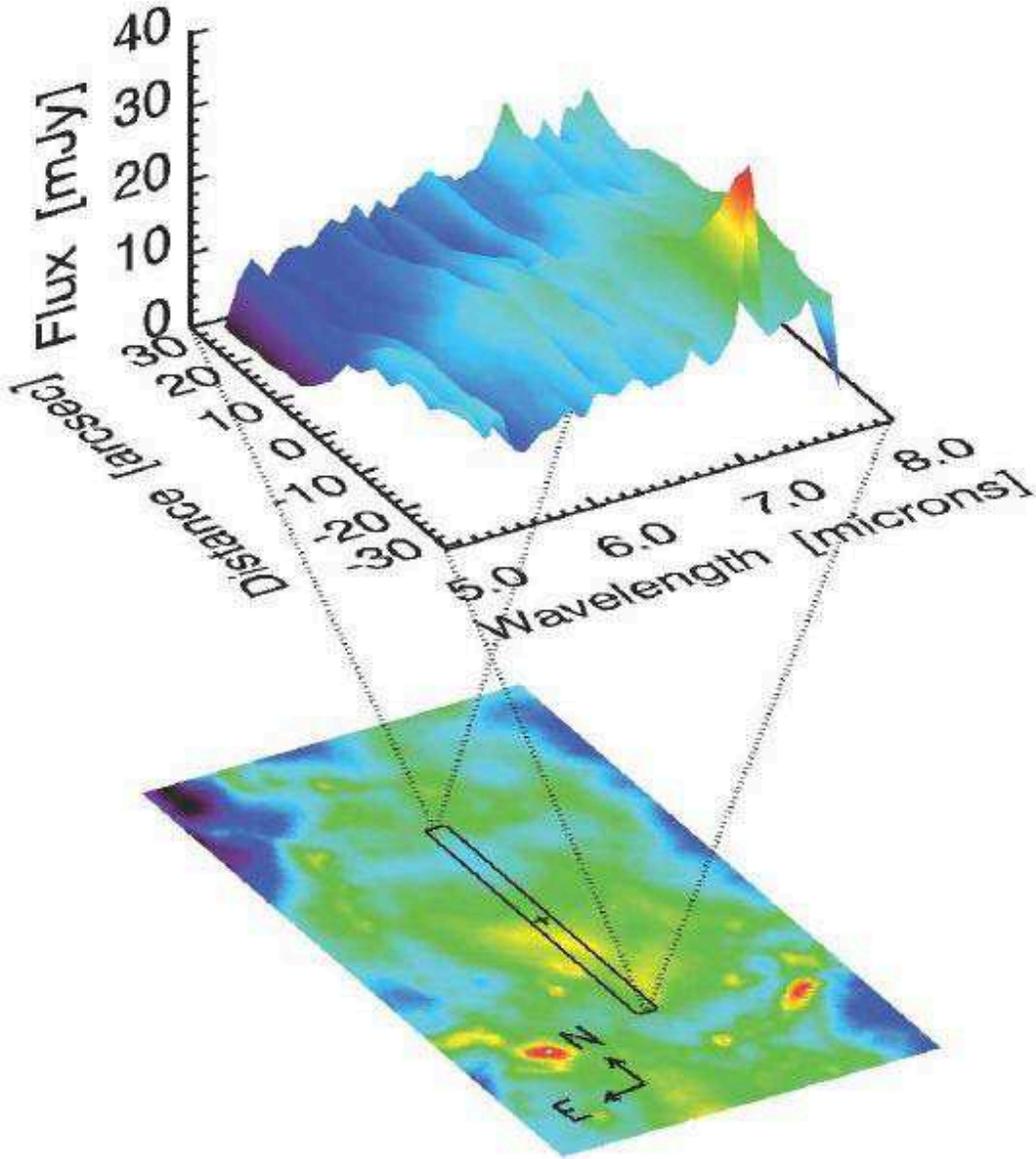} \caption{\label{TLR-4} The IRS
Short-Low second-order spectrum and the IRAC 8 $\micron$ image. In
this figure the location of the IRS Short-Low second-order aperture
is indicated on the IRAC image in the lower part of the figure, with
the location of the IRS-Tgt-Cntr position is indicated by a small
cross. The spectral dispersion of this spatial cut through the image
is shown in the upper part of the figure. The poor correlation of
the [ArII] line at 7.0 $\micron$ with the IRAC structure is
immediately obvious.}
\end{figure}

\begin{figure}
\epsscale{1.0}
\plotone{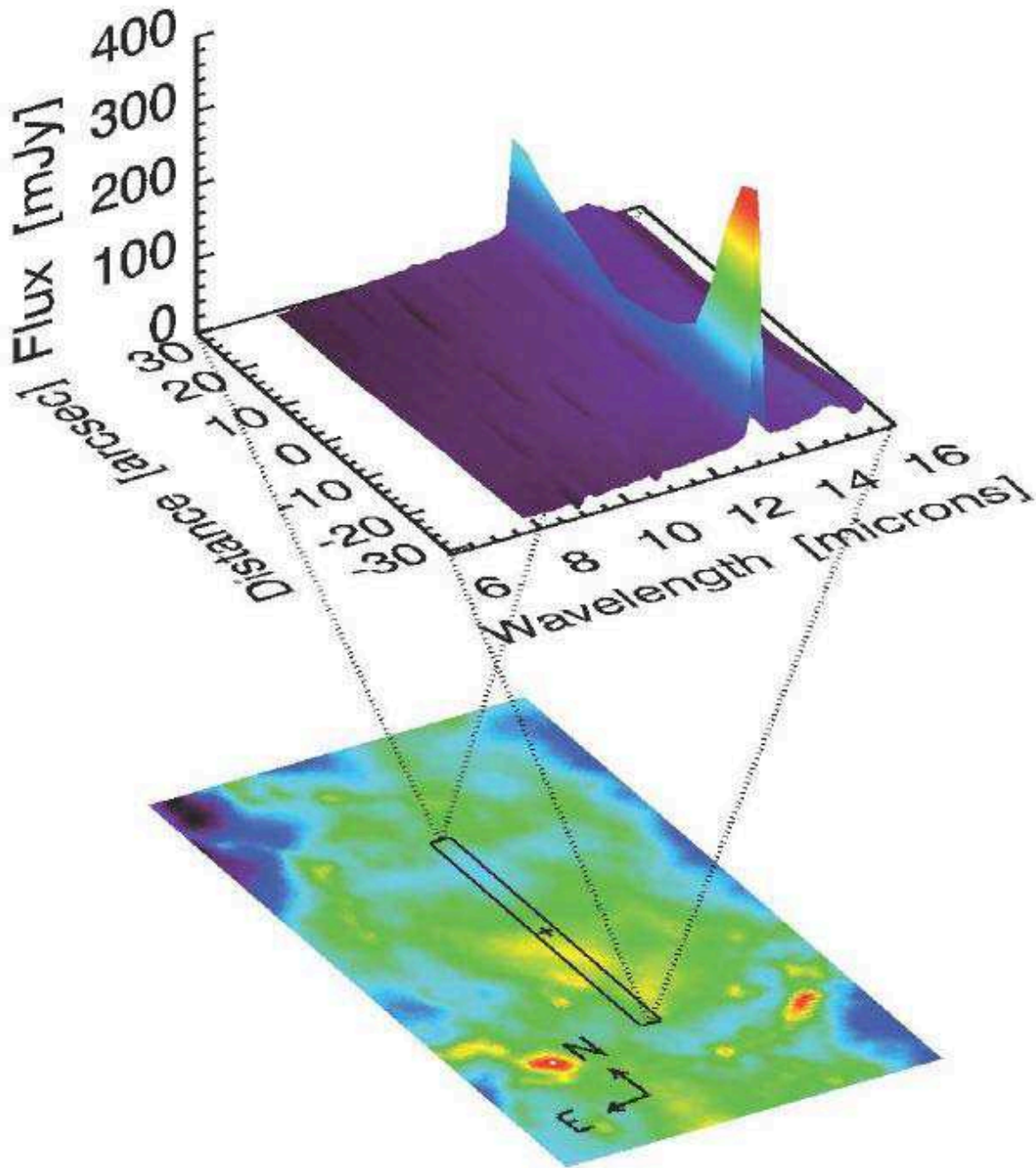}
\caption{\label{TLR-5} The IRS
Short-Low first-order spectrum and the IRAC 8 \micron{} image. This
figure is the same as Figure \ref{TLR-4}, except that it displays
the IRS first-order spectrum. The slit location on the IRAC image is
the same for both Short-Low orders. In this case the strong line of
[NeII] at 12.8 \micron{} is obvious.}
\end{figure}

With the long IRS slits in the low-resolution modules it is possible
to observe variations in the emission line strengths over the
different regions within the nebula sampled by the slits. In
general, the IRS long-slit spectra show a relatively spatially
smooth synchrotron continuum with much more pronounced spatial
structure in the emission line fluxes. These line emission spatial
variations can be quite strong and are shown in Figures \ref{TLR-4}
and \ref{TLR-5}, which show the correlation between the emission
line strength and the observed IRAC 8 $\micron$ spatial structure.
It is tempting to infer that the IRAC 8 $\micron$ image spatial
structure is due to the 7.0 $\micron$ [ArII] line emission. As can
be seen in the figures, there definitely is some spatial
correlation, but compared to the integrated synchrotron emission
within the IRAC band the flux in the [ArII] line is relatively low
and the variation in the line intensity by itself would not be
enough to account for all the observed 8 $\micron$ structure.
Emission lines at other wavelengths are not necessarily correlated
with the structure observed by IRAC. For example Figure \ref{TLR-5}
shows that although the 12.8 $\micron$ [NeII] line emission has a
peak in the region of strong IRAC 8 $\micron$ emission, it also
rises in strength in an 8 $\micron$ ``hole''. In addition, the
region of strongest IRAC emission corresponds to the lowest [NeII]
emission.

\subsection{IRS Spectra \label{irs_spec}}

\begin{figure}
\epsscale{1.0}
\plotone{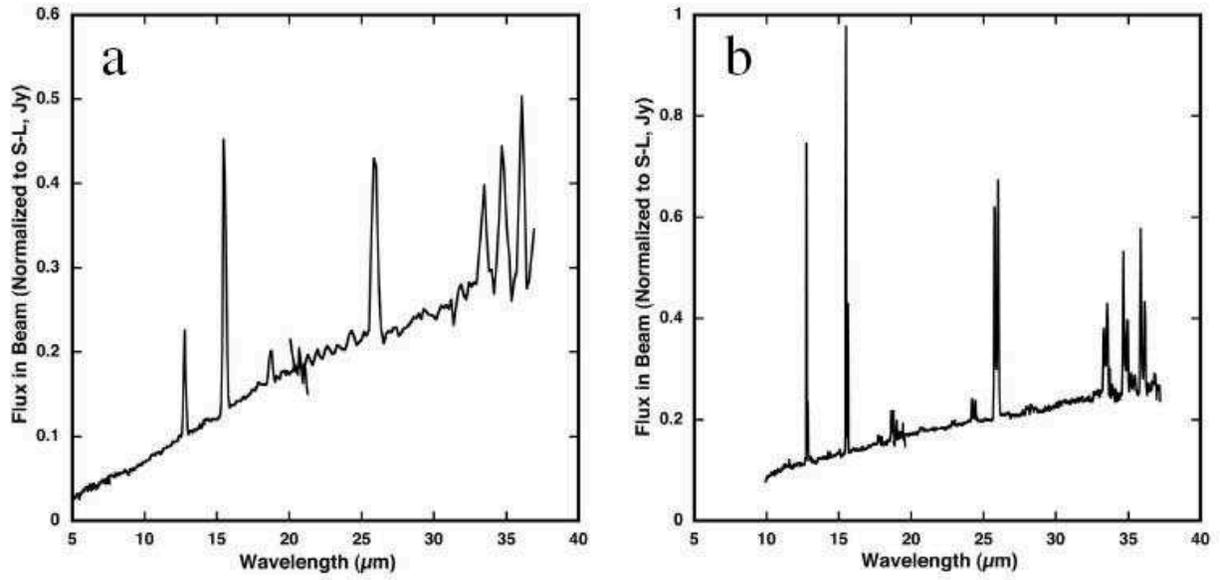}
\caption{\label{TLR-3} The complete low and high resolution spectra of
the Crab nebula at the IRS-Tgt-Cntr location. The fluxes were scaled to
the Short-Low module data as described in the text to correct for the different
aperture dimensions (\S \ref{irs_spec}).}
\end{figure}

The IRS modules' slits are of different sizes and
had different orientations on the Crab (see Fig. \ref{TLR-1}).
In order to compare the spectra for the different IRS modules, we have
used the smoothness of the synchrotron continuum emission as observed
in the IRAC images to calculate scaling factors -- in the analysis
below the IRS observed data were scaled so that their synchrotron continua were
in agreement with the levels seen from the Short-Low module. The
scaling factors that were employed to achieve this agreement were
0.24 for the Long-Low module, 1.0 for the Short-High module, and
0.33 for the Long-High module. These scaling factors are all larger
by approximately a factor of two compared to the ratio of the beam
sizes, which indicates that the nebular emission is not completely smooth over
these spatial dimensions. Figure \ref{TLR-3} shows the low and
high-resolution spectra after this scaling, with the short and long
modules' data combined. The reasons for the slight difference
between the synchrotron continuum slopes in the high and
low-resolution modules is due to residual uncorrected diffraction
effects. The IRS data from the SSC data pipeline is flux-calibrated
with point sources. Although corrections have been applied to
account for the relatively uniform extent of the synchrotron
emission compared to the point-source calibrators, this correction
is not perfect.

The fine structure lines observed with the IRS high-resolution
modules were fit with Gaussian curves and converted into line
strengths.  The resultant line identifications, line heights, and line
fluxes are given in Tables \ref{sh} and \ref{lh}. As mentioned above,
the synchrotron baseline was used to renormalize the Long-High
module line strengths to correspond to the Short-High module.

The observed spectra show a steadily-rising continuum due primarily
to synchrotron emission, and to a lesser extent to the zodiacal
background, with various fine-structure emission lines superimposed.
As was discussed above, the strength of the fine structure emission
lines varies strongly with position within the nebula -- for
example, the spectrum at the IRS-Tgt-Cntr position does not show the
7.0 $\micron$ [ArII] emission line that is seen strongly in other
positions in the nebula. In the high-resolution spectra the emission
lines are observed to be split due to Doppler shifts in the front
and back sides of the expansion envelope. The average
$\Delta\lambda$/$\lambda$ difference between the front and back
shell emission line wavelengths is 0.00843, which corresponds to a
shell expansion velocity of 1264 km s$^{-1}$.

The IRS fine structure emission lines show evidence of high
excitation that may be caused by photoionization by synchrotron radiation. 
For example, we observe emission from [NeV] which has an ionization
energy of 97 eV. Given the violence of the Crab expansion, shocks
sufficient to excite forbidden lines are also not unexpected. 
With the large number of fine structure emission lines observed
by the IRS, it should be possible in principle to derive electron
temperatures and densities from the line ratios. Unfortunately,
for all of the lines coming from the same ion, the line pairs
appear in different IRS modules. Modules' different
aperture sizes, different slit position angles, coupled
with the observed spatial structure in the emission line
emission regions, result in line ratios that have large and unquantifiable
uncertainties. For example, in comparing the measured 15.5
$\micron$/36 $\micron$ line ratio for [NeIII], we find ratios of 3.3
$\pm 0.3$ for the red-shifted lines, and 5.6 $\pm 0.3$ for the
blue-shifted lines. For temperatures around 10,000K the predicted
ratio for these lines ranges from 11 to 38 for electron number
densities ranging from 10 to 10$^{7}$ cm$^{-3}$. Similarly for the
18.7 $\micron$/ 33.4 $\micron$ line ratio for [SIII] we find that
the measured values for both the red and blue lines are less than
the predicted level even for the lowest electron densities. This
indicates that there is more line flux entering the Long-High IRS aperture
than is expected, even after normalization using the synchrotron
continuum baseline. Given the spatial structure that can be seen in
the images of the IRS high-resolution module slits this is not
surprising -- the larger Long-High slit must contain more bright emission
knots than does the Short-High slit.

\subsection{Line emission or Dust in the MIPS 24~$\micron$ Image? \label{line_or_dust}}

\begin{figure}
\epsscale{1.0}
\plotone{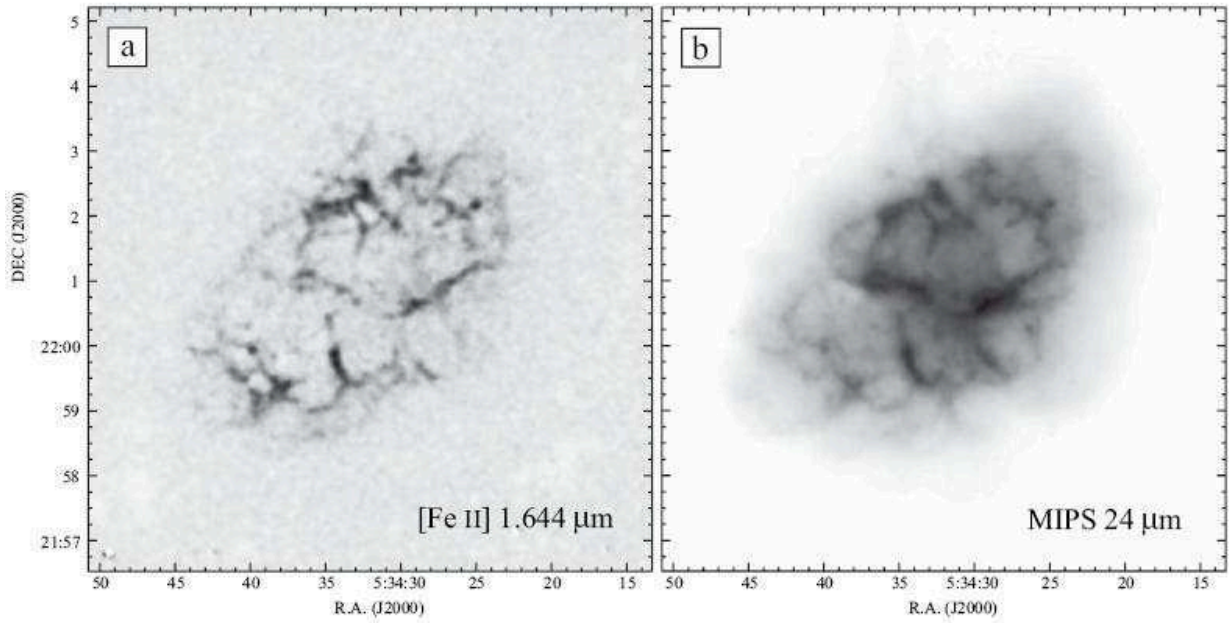}
\caption{\label{Fe2comp} Comparison
of the visual image made in the light of [Fe II] in panel (a) with
MIPS 24 $\micron$ image in panel (b).}
\end{figure}

IRS spectra of the Crab discussed above show that a bright emission
feature --- either [Fe~{\sc ii}] 25.99~$\micron$ or [O~{\sc iv}]
25.89 $\micron$ --- contributes significantly to the MIPS
24~$\micron$ image. Both lines have been observed in young supernova
remnants and the Crab \citep[e.g.,][]{Are99, Oli99, Gre04}. In Figure
\ref{Fe2comp} we compare the MIPS 24~$\micron$ image to our
ground-based near-IR image of the Crab taken through a narrow filter
that isolates [Fe~{\sc ii}] $a^4F_{9/2}-a^4D_{7/2}$ at
1.644~$\micron$, with continuum emission subtracted. The major
difference between the two images in Figure \ref{Fe2comp} is in the
filaments that run east to west across the middle of the remnant;
while these filaments appear thin and rather faint compared to some
of the brighter [Fe~{\sc ii}] 1.644~$\micron$ filaments in the north
and south of the remnant in Figure \ref{Fe2comp}$a$, they are by far
the brightest emission features in the 24~$\micron$ MIPS image in
Figure \ref{Fe2comp}$b$. Why are these filaments so much brighter
in the MIPS 24~$\micron$ image?

An issue of potential concern is whether the narrow bandpass of the
1.644~$\micron$ filter cuts-out some of the emission from fast
moving filaments that may be Doppler shifted to extreme velocities,
because this could cause spurious differences between the MIPS image
and the [Fe~{\sc ii}] 1.644~$\micron$ emission map.  However, the
observed velocities from the brightest filaments are actually at
relatively low velocities in the pinched waist of the Crab. The
Doppler shifts are in the range of $\pm$500 to 1200 km s$^{-1}$,
with most filaments moving at $\pm$800 to 900 km s$^{-1}$ (Smith
2003). These velocities are well within the 1\% ($\pm$1500 km
s$^{-1}$) FWHM bandpass of the [Fe~{\sc ii}] filter. Even if some
of the fainter and faster filaments are lost due to their Doppler
shifts, it is reassuring that the relative brightness distribution
of the filaments is similar to that in previous [Fe~{\sc ii}] and
other IR emission-line images of the Crab \citep[e.g.,][]{Hes90,
Gra90}.

Thus, the striking differences in brightness between the various
filaments of the Crab in the [Fe~{\sc ii}] 1.644~$\micron$ and
24~$\micron$ MIPS image are real, and are due either to 1) extreme
differences in physical conditions that cause very different
relative strengths of [Fe~{\sc ii}] 25.99$\micron$ and
1.644~$\micron$ in certain filaments, 2) differences in ionization
level or chemical abundances that cause stronger [O~{\sc iv}] 25.89
$\micron$ or weaker [Fe~{\sc ii}] emission there, or 3)
substantially more warm dust in those filaments.

The flux ratio of [Fe~{\sc ii}] 1.644~$\micron$/25.99~$\micron$
depends on both electron density and temperature, in the sense that
the ratio tends to be larger for higher densities and especially for
higher temperatures. The relative strengths of IR lines of [Fe~{\sc
ii}] can be predicted by quantitative models like those described by
\citet{Har04}. In the Crab's filaments, typical electron densities
and temperatures derived from visual wavelength diagnostics are $n_e
\simeq 1300$ cm$^{-3}$ and $T_e \simeq 11,000-18,000$ K
\citep[e.g.,][]{Fes82, Mac89, Dav85}. Under these conditions, one
would expect [Fe~{\sc ii}] 1.644~$\micron$ to be roughly 3--4 times
stronger than the 25.99~$\micron$ line. From our flux-calibrated
narrowband image, we measure a total [Fe~{\sc ii}] 1.644~$\micron$
flux for the whole Crab Nebula of 6.3($\pm$1.5)$\times$10$^{-11}$
ergs s$^{-1}$ cm$^{-2}$. If the average physical conditions quoted
above dominate throughout the filaments, the total 25.99~$\micron$
flux should be only 1.5--2$\times$10$^{-11}$ ergs s$^{-1}$
cm$^{-2}$. The total observed MIPS 24~$\micron$ flux density is
about 60 Jy ($\sim$3$\times$10$^{-10}$ erg s$^{-1}$ cm$^{-2}$
$\micron^{-1}$), and is clearly dominated by emission from the
filaments, not the synchrotron continuum (Fig. \ref{Fe2comp}$b$).
Integrated over the MIPS filter bandpass, the total flux included in
the 24~$\micron$ filter is then $\sim$1.5$\times$10$^{-9}$ erg
s$^{-1}$ cm$^{-2}$. Thus, [Fe~{\sc ii}] 25.99~$\micron$ should not
make much of a contribution to the MIPS 24~$\micron$ image (only
about 1\% of the total flux included in that filter).

On the other hand, if the filaments have cores that are shielded
from the UV radiation field of the synchrotron nebula, then the
prevailing electron temperatures traced by visual-wavelength
diagnostics might not apply, and the expected [Fe~{\sc ii}]
1.644~$\micron$/25.99~$\micron$ ratio could be very different. For
example, if the electron temperature were as low as 1,000--2,000 K
in the cores of some filaments, then at the same density of
$\sim$1000 cm$^{-3}$ the 25.99~$\micron$ line could be more than
1000 times stronger than [Fe~{\sc ii}] 1.644~$\micron$. However,
under these same conditions, [Fe~{\sc ii}] 5.430~$\micron$ would be
roughly 800 times brighter than [Fe~{\sc ii}] 1.644~$\micron$, and
almost as strong as the 25.99~$\micron$ line. Based on the rather
weak 5.34~$\micron$ line seen in IRS spectra, and the fact that the
``equatorial'' filaments are not conspicuous in the IRAC
5.8~$\micron$ image, it would seem that the filaments do not contain
shielded cool cores at very low temperatures, and therefore, that
the majority of the MIPS 24~$\micron$ emission is not [Fe~{\sc ii}].

Instead, the bright emission feature in the MIPS 24~$\micron$ filter
is most likely dominated by the [O~{\sc iv}] 25.89~$\micron$ line.
This agrees with our analysis of the IRS spectra, suggesting that
those conclusions are valid across the remnant in areas not covered
by the IRS aperture. In visual wavelength spectra, the [O~{\sc
iii}] $\lambda$5007 line is particularly bright in the equatorial
filaments, compared to other emission lines \citep{Smi03, Mac89}.
Thus, it stands to reason that the much brighter emission from
equatorial filaments in the MIPS 24~$\micron$ image is due to
enhanced [O~{\sc iv}] there as well, either because of higher
ionization or enhanced oxygen abundances compared to the other
filaments in the Crab Nebula.

Another possibility is that emission from large dust grains in the
filaments contributes to the excess equatorial emission seen by
MIPS. \citet{Gre04} found that the gas:dust mass ratio in the Crab
was comparable to the normal interstellar value, with an upper limit
to the dust mass of $\sim$0.2 M$_{\odot}$. However, the presence of
strong line emission in the MIPS 24~$\micron$ image makes it
difficult to put reliable constraints on the amount of dust, if any,
in the Crab Nebula. The MIPS 70~$\micron$ image also shows enhanced
emission from the same equatorial filaments, which once again, could
result from a contribution of either warm dust or [O~{\sc iii}]
88.36~$\micron$ emission. However, \citet{Gre04} found that the
ISO/LWS spectrum of the Crab was dominated by continuum from dust
and synchrotron, not line emission. Thus, the apparent excess far-IR
emission above the synchrotron continuum level in the integrated
flux density of the Crab (Fig. \ref{nebulaflux}) may be due to
either dust or line emission.

Whether the bright MIPS emission is caused by excess dust or by
stronger high-excitation emission lines, the comparison of the
[Fe~{\sc ii}] 1.644~$\micron$ image and the MIPS 24~$\micron$ image
reinforces the long-held notion that there is something very
different about the ``equatorial'' filaments that run east--west
across the middle of the Crab Nebula. These are the same filaments
that are known to have different chemical abundances than the rest
of the Crab -- most notably a higher He/H abundance
\citep[e.g.][]{Uom87, Fes82}. One hypothesis is that the supernova
exploded into a pre-existing circumstellar ring, much like the ring
around SN~1987A \citep{Sug02} or the ring that is thought to have
been swept up by the Crab-like SNR~0540-69.3 in the LMC
\citep{Mor06}. Indeed, the filaments that bisect the Crab seem to
connect the dark bays of the synchrotron nebula, which were
presumably formed as the expanding plerion was pinched by an
equatorial disk or ring \citep{Fes92}.  Additionally, the global
kinematics of the Crab's shell, traced best in [O~{\sc iii}]
emission, seem to be bipolar, with a pinched waist coinciding with
these ``equatorial'' filaments \citep{Mac89, Smi03}. Finally, the
mysterious northern ``jet'' or ``chimney'' \citep{Che75, Gul82} runs
along an axis perpendicular to this putative equatorial ring. Taken
together, these clues may suggest that the Crab's progenitor star
had a rotation axis oriented roughly north/south, and that the axis
changed dramatically during the supernova explosion to match the
currently-observed axis of the pulsar's jet at about --45$\arcdeg$.

\subsection{Estimate of Dust Mass \label{dust_estimate}}

The presence of the bump between 24 $\micron$ and 100 $\micron$ in
the integrated flux density of the Crab Nebula may suggest that
there is a small amount of dust present in the nebula (see Figure
\ref{nebulaflux}). In order to estimate a rough upper limit on the
dust mass that may account for this infrared excess, we attempted to
remove the contributions from line emission and synchrotron
continuum.

\citet{Gre04} found
that the excess in the ISOPHOT fluxes between 60 $\micron$ and 100
$\micron$ can be explained by a small amount of warm dust
corresponding to 0.01--0.07 M$_{\odot}$ of silicates at 45 K or
0.003--0.02 M$_{\odot}$ of graphite at 50 K. We followed the method
of \citet{Gre04} and subtracted the extrapolated synchrotron
continuum from the MIPS 24 $\micron$ and 70 $\micron$ fluxes based
on the synchrotron power law 25.9($\nu$/18.7 Thz)$^{-0.5}$ given by
\citet{Gre04} and derived by \citet{Dou01} from the ISOCAM
mid-infrared spectro-imaging observations. Based on the ISO spectra,
\citet{Gre04} also found that the line emission contribution to the
60 and 100 $\micron$ fluxes is 8 \% and 7 \%, respectively. Since
ISO spectra show no additional emission lines centered at 70
$\micron$, we approximate that the upper limit on the contribution
from line emission to the 70 $\micron$ integrated flux is on the
order of 7 \%, corresponding to $\sim$ 11 Jy. The main difficulty
arises in estimating the contribution from line emission to the 24
$\micron$ flux. By integrating the IRS spectrum over the MIPS 24
$\micron$ bandpass, we find that the contribution from line emission
is approximately 5 \% of the total flux in the beam. For the purpose
of estimating a rough upper limit on the dust mass, we applied this ratio
to the total integrated flux at 24 $\micron$, which gives a line
contribution of $\sim$ 2.8 Jy. Since the strength of line emission
in the Crab has a strong spatial dependence and peaks along the filaments,
we expect that this assumption significantly underestimates the
amount of line emission at 24 $\micron$. The IRS spectrum used
for the estimate is centered on a region near the center of the
nebula that does not include any filaments.

After subtracting the synchrotron continuum and line emission from
the MIPS 24 $\micron$ and 70 $\micron$ data, and ISOPHOT 60 $\micron$
and 100 $\micron$ data from \citet{Gre04}, we fitted a blackbody
distribution to the residual flux densities. The resulting blackbody
distribution peaks at a temperature of $\sim$ 74 K, with
($\lambda$F$_{\lambda}$)$_{max}$ $\sim$ 6.28$\times$10$^{-9}$ erg
s$^{-1}$ cm$^{-2}$. We expect that these values are highly uncertain due
to the sensitivity of the blackbody fit to the 24 $\micron$ data
point. We calculate a dust mass using the following relation

\begin{equation}
M_{d}=\frac{1.36({\lambda}F_{\lambda})_{max} a \rho
4{\pi}d^2}{3{\sigma}T_{d}^4}
\end{equation}

\noindent where $a$ is the grain size, $\rho$ is the grain density, $d$ is the
distance to the Crab Nebula, and $T_d$ is the dust temperature
\citep{Geh98}. Assuming a grain size of 10 $\micron$, graphite
grains ($\rho$ = 2.25 gm cm$^{-3}$), and a distance of 2.1 kpc, we
derive a dust mass of $\sim$ 0.001 M$_{\odot}$. We emphasize that
due to the unknown line contribution to the 24 $\micron$ flux, this
number is only a rough estimate of the dust mass. If we assume that
the excess emission at 24 $\micron$ is dominated by line emission
and that the blackbody peaks at the 70 $\micron$ data point, we find
a dust mass of 0.004--0.010 M$_{\odot}$ for graphite grains and
0.006--0.015 M$_{\odot}$ for silicates.

\subsection{Synchrotron Spectra \label{synch_spec}}

IRAC 3.6 $\micron$ and 4.5 $\micron$ images are dominated, for the
Crab, by the synchrotron emission that has been observed from radio
through X-ray frequencies, with only little evidence for the
filamentary line emission seen, e.g., in IRAC 8 $\micron$ or in the
MIPS 24~$\micron$ images. The very hard (flat) radio spectra in
plerions (i.e., filled-center supernova remnants) like the Crab imply
correspondingly hard relativistic electron distributions;  the
origins of these are still unknown \citep{node04}. Studies of the
variation of the spectral index, $\alpha_{\lambda}(RA, Dec)$ i.e.,
the local slope of the synchrotron spectrum, as a function of both
frequency and position in the remnant, provide the key data
necessary to understand both relativistic particle acceleration and
loss processes.

\begin{figure}
\epsscale{0.8} \plotone{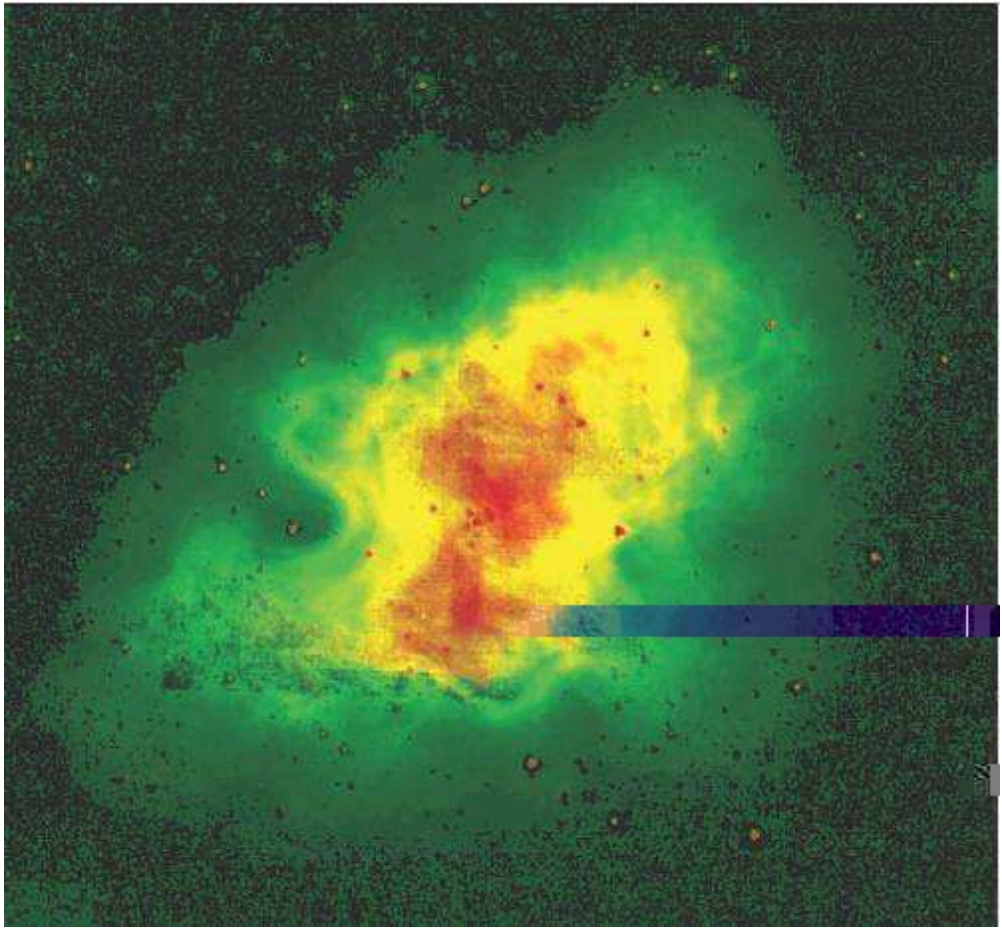} \caption{\label{SynchColor} The
synchrotron spectral index distribution is shown in color, with the
local intensity from the IRAC 3.6 $\micron$ image. Note the flat
spectrum emission from the torus and jets extends far from the
pulsar. Red (green) corresponds to spectral indices of 0.3 (0.8).}
\end{figure}

We began our examination of the spectral index as a function of
position using IRAC 3.6 $\micron$ and 4.5 $\micron$ images, as shown
in Figure \ref{SynchColor}. Note that the emission associated with
the jet and torus is flat (red) and extends far from the pulsar with
little or no steepening visible. The underlying nebular emission
progressively steepens with distance. This steepening away from the
jet and torus is also seen in the power law indices derived by
\citet{willing} from XMM-Newton observations around 2 keV. At present, it
is not possible to determine whether this steepening is with
distance from the pulsar or distance from the jet; this distinction
is physically important for understanding the acceleration and
transport of relativistic particles. The absolute values of the
spectral indices would be uncertain by approximately $\pm 0.3$ if we
estimate that the calibration uncertainties in each of the IRAC
bands are approximately 5\%. However, the integrated fluxes between
3.6 $\micron$ and 4.5 $\micron$ yield a spectral index similar to
the derived value of 0.5 by \citet{Dou01}, and therefore we do not
include calibration errors in the discussion below. Any
uncertainties in the overall spectral index do not affect the
spatial variations in the indices as seen in Figure
\ref{SynchColor}.

We next looked at the overall shape of the synchrotron spectrum by
examining its mapping into ``color-color'' space, using the
technique described by \citet{Kat93}.  In this method, we add
information at optical wavelengths, and then plot $\alpha_{IR}(RA,
Dec)$ vs. $\alpha_{opt}(RA,Dec)$ for all positions in the Crab. This
plot is then compared to various fiducial spectral shapes to
determine the underlying broadband shape of the synchrotron
spectrum. This method provides a powerful way to map out the
synchrotron spectrum in detail, although measurements are available
at only a small number of wavelengths. It is based on the {\em
ansatz} that the {\em shape} of the relativistic electron
distribution is constant throughout the nebula; spatial variations
in magnetic field strength and in the degree of adiabatic and
radiative losses then lead to spatial variations in the observed
spectral index.

\begin{figure}
\epsscale{1.0}
\plotone{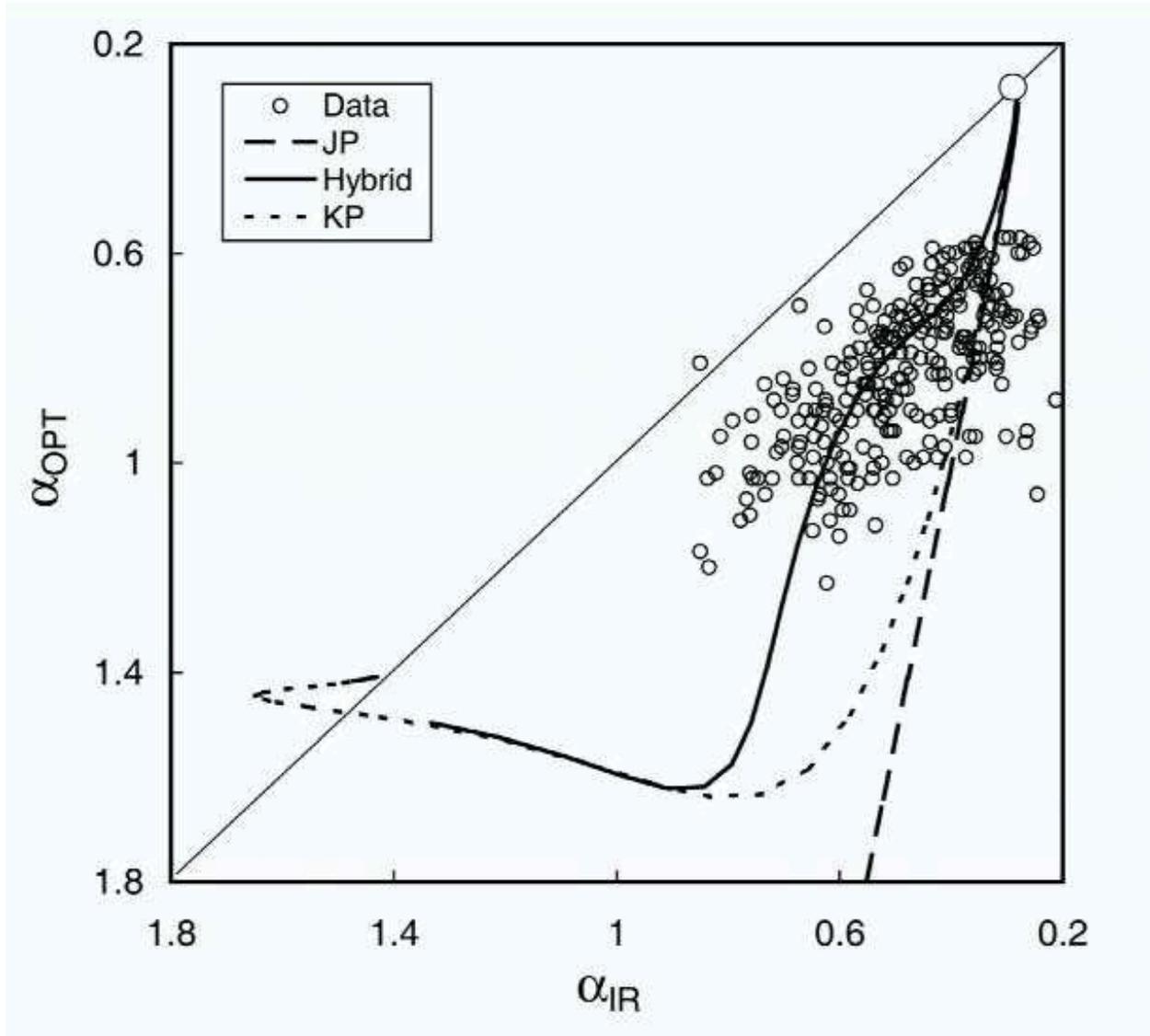}
\caption{\label{ccIRopt} Color-color
diagram, representing spectral indices in 10\arcsec boxes in the
infrared, and between the infrared and optical, as discussed in the
text (\S \ref{synch_spec}).}
\end{figure}

Our first ``color-color'' diagram is shown in Figure \ref{ccIRopt},
where we have used the visual spectral indices from \citet{Ver93},
who calculated the mean spectral index in 10\arcsec boxes between
9241 \AA\ and 5364 \AA\ to compare with equivalent 10$\arcsec$ box
spectra calculated from IRAC 3.6 $\micron$ and 4.5 $\micron$ images.
The IRAC fluxes were corrected for extinction using the interstellar
extinction values from \citet{Rie85} and an adopted value for A$_V$
of 1.5. The combination of statistical errors of approximately 1\%
in each band and variations in inhomogeneities between different
lines of sight together are responsible for the scatter of the data
around the best fit spectral shape.

Comparison of the ``color-color'' data with fiducial homogeneous
spectral models shows that these models are not adequate in describing
the data. Two standard spectral shapes are shown for comparison -- the JP \citep{Jaf73}
model in which relativistic electrons are continuously isotropized
in pitch angle, leading to an exponentially cutoff spectrum at high
energies, and the KP \citep{Kar62} model where losses lead to an
anisotropic pitch angle distribution, and a high frequency power
law. Each spectral model is anchored to the 0.3 spectral index
observed throughout the nebula at radio frequencies \citep{Swi80}.
Here, it can be seen that both the JP exponential cutoff, and KP
spectral shapes are ruled out by the relationship between the
optical and IR fluxes. This conclusion is not affected by the
calibration errors discussed above; these simply result in a uniform
translation of the data in the color-color plane, and would still
not agree with the models.

Although a completely new relativistic electron distribution could
be derived to fit the data, a much simpler alternative is to explore
simple combinations of the standard spectra. We show one such
combination in Figure \ref{ccIRopt}, a ``hybrid'' model representing
the presence of two KP spectra at each position. The two spectra
shown here have cutoff frequencies (low frequency brightnesses) in
the ratios of $3:1 (1:1.8)$.  These particulars are not relevant,
since there are a variety of spectral combinations, with different
numbers of components, cutoff frequencies, normalizations, etc.,
that could equally well fit the data. However, they all point to
some inhomogeneities along the line of sight which smear out the
spectrum (reduce its curvature) in the optical-IR regime.

In the Crab, the inhomogeneity indicated by the spectral shape
likely arises from the superposition of the jet and torus onto the
larger nebula. This can be be seen by subtracting IRAC 3.6 $\micron$
image from the 5.8 $\micron$ image, with the resulting emission seen
in Figure \ref{residsynch}. The most striking thing in this image is
the absence of the torus and jet regions that dominate the emission
in both bands. This is due to the flatter nature of those regions,
which extend over large distances in the nebula. Thus, our data are
consistent with emission that shows little or no spectral steepening
with distance from the pulsar, superposed on the rest of the nebula,
which does steepen. The existence of these two components was
suggested by \citet{Ban02} in their comparison between 20 cm and 1.3
mm images. \citet{Gre04} disagreed with these conclusions based on
their SCUBA 850 $\micron$ images, but their data also show excess
flattening due to superposed components. The fact that the torus and
jet remain flat to higher frequencies than the rest of the nebula
makes them much more prominent in IR, optical and X-ray images,
compared to those in the radio \citep{Bie04}. It also means that
sophisticated models will be necessary to understand the
acceleration and transport of relativistic particles throughout the
nebula.

\begin{figure}
\epsscale{1.0} \plotone{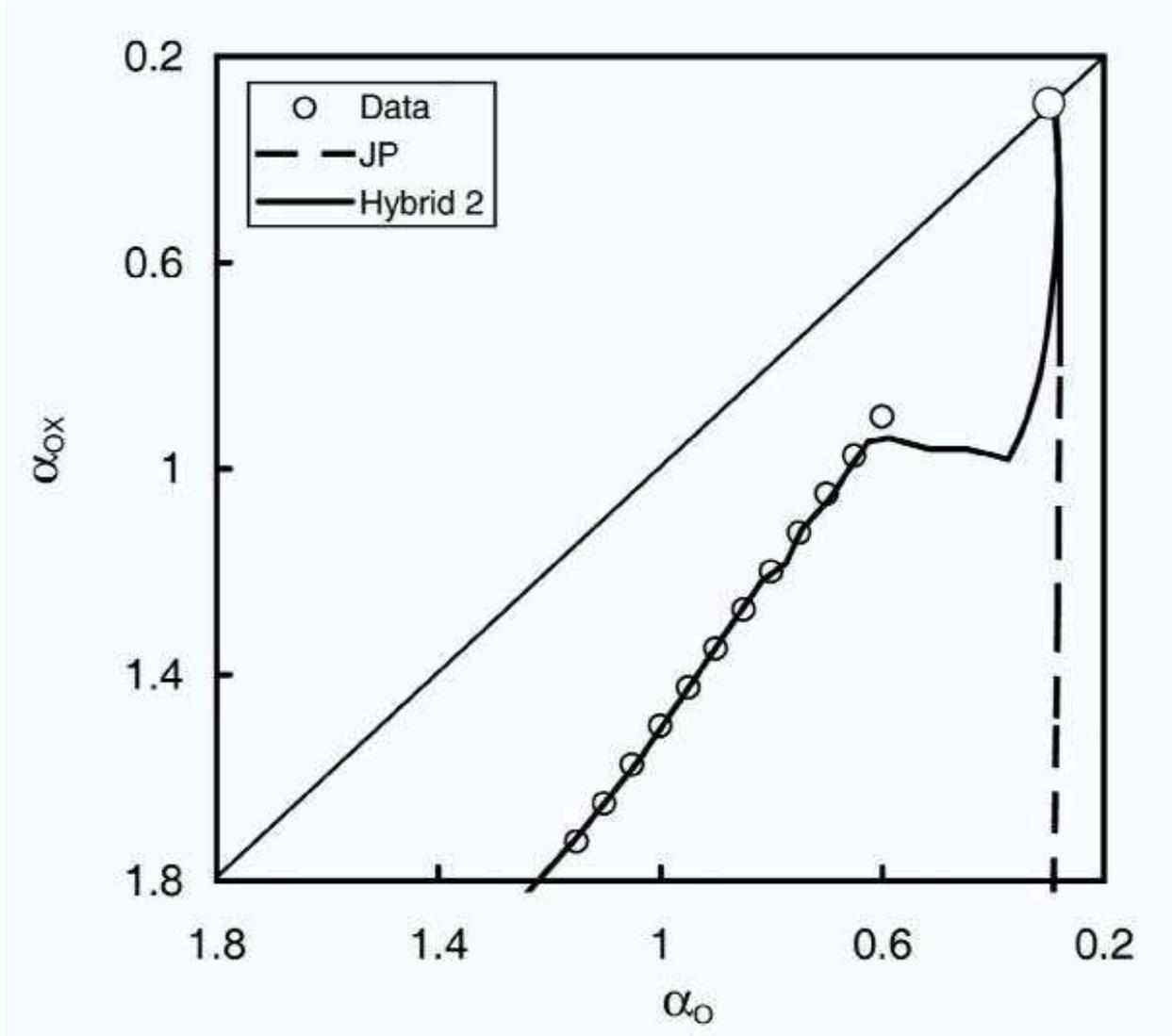}
\caption{\label{ccoptX} Color-color diagram comparing the optical spectral indices
to those in the X-ray, as described in the text (\S \ref{synch_spec}).}
\end{figure}

To explore the higher energy end of the spectrum, we used a second
color-color diagram, now relating the optical spectral indices to
those between the optical and X-ray, as presented by \citet{Ban98}.
For the data, we took the trend line between $0.8 < \alpha_{opt-X} <
1.75$ which can be represented as $\alpha_{opt-X} = 1.5 \times
\alpha_{opt}$. We consider the very weak X-ray data with
$\alpha_{opt-X} > 1.75$ (above line (c) in their Figure 3) to be
unreliable because it shows no correlation with the optical spectral
indices. Note that the detectable X-ray synchrotron nebula is
considerably smaller than at lower frequencies. The $\alpha_{opt}$
vs. $\alpha_{opt-X}$ data line (represented by a series of open
circles) and spectral shape models are shown in Figure \ref{ccoptX}.

\begin{figure}
\epsscale{1.0}
\plotone{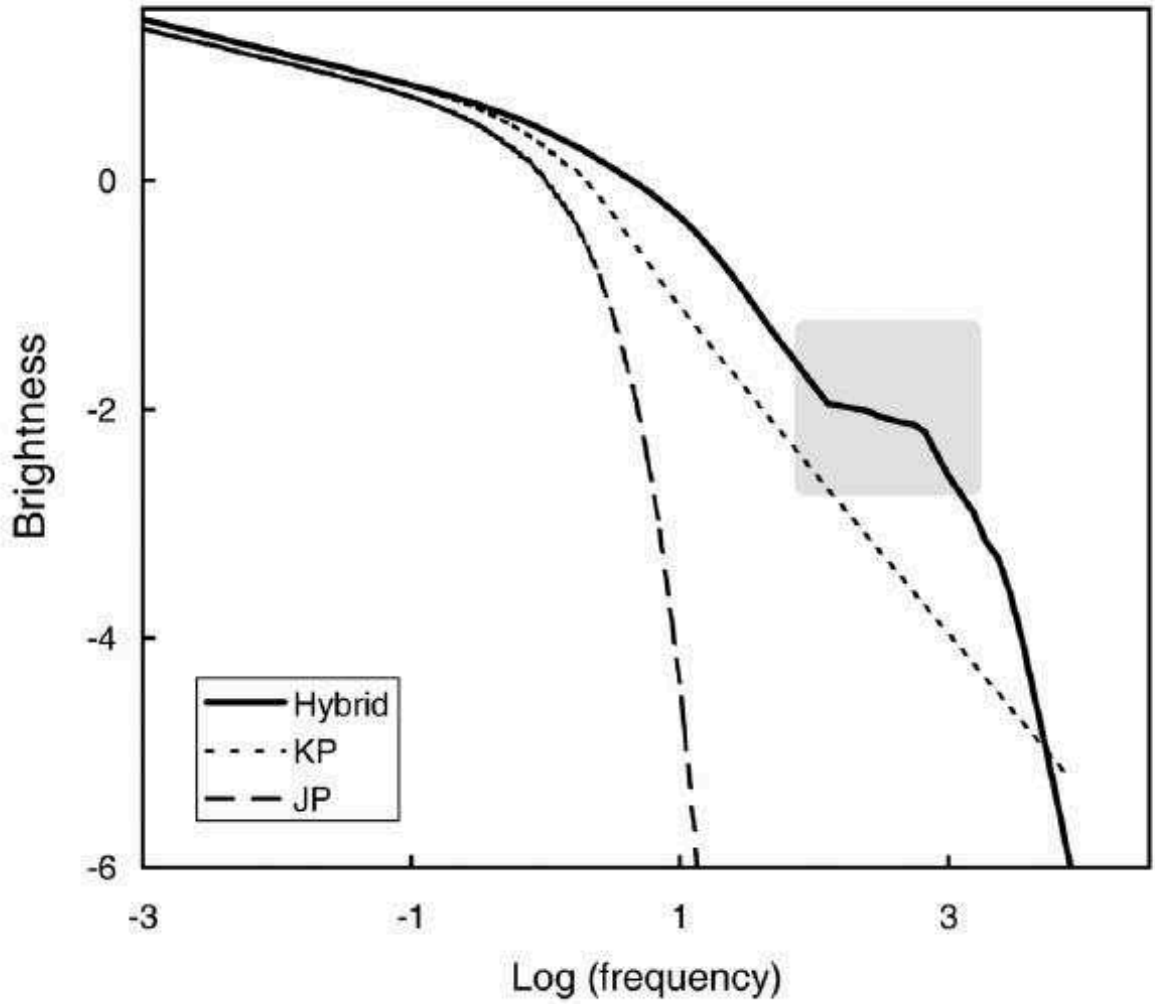}
\caption{\label{crabspec2} The derived shape
for the broadband synchrotron spectrum of the Crab, as shown in
color-color space in the previous two figures. The shaded box
indicates where the details of the shape are not determined.}
\end{figure}

\begin{figure}
\epsscale{1.0} \plotone{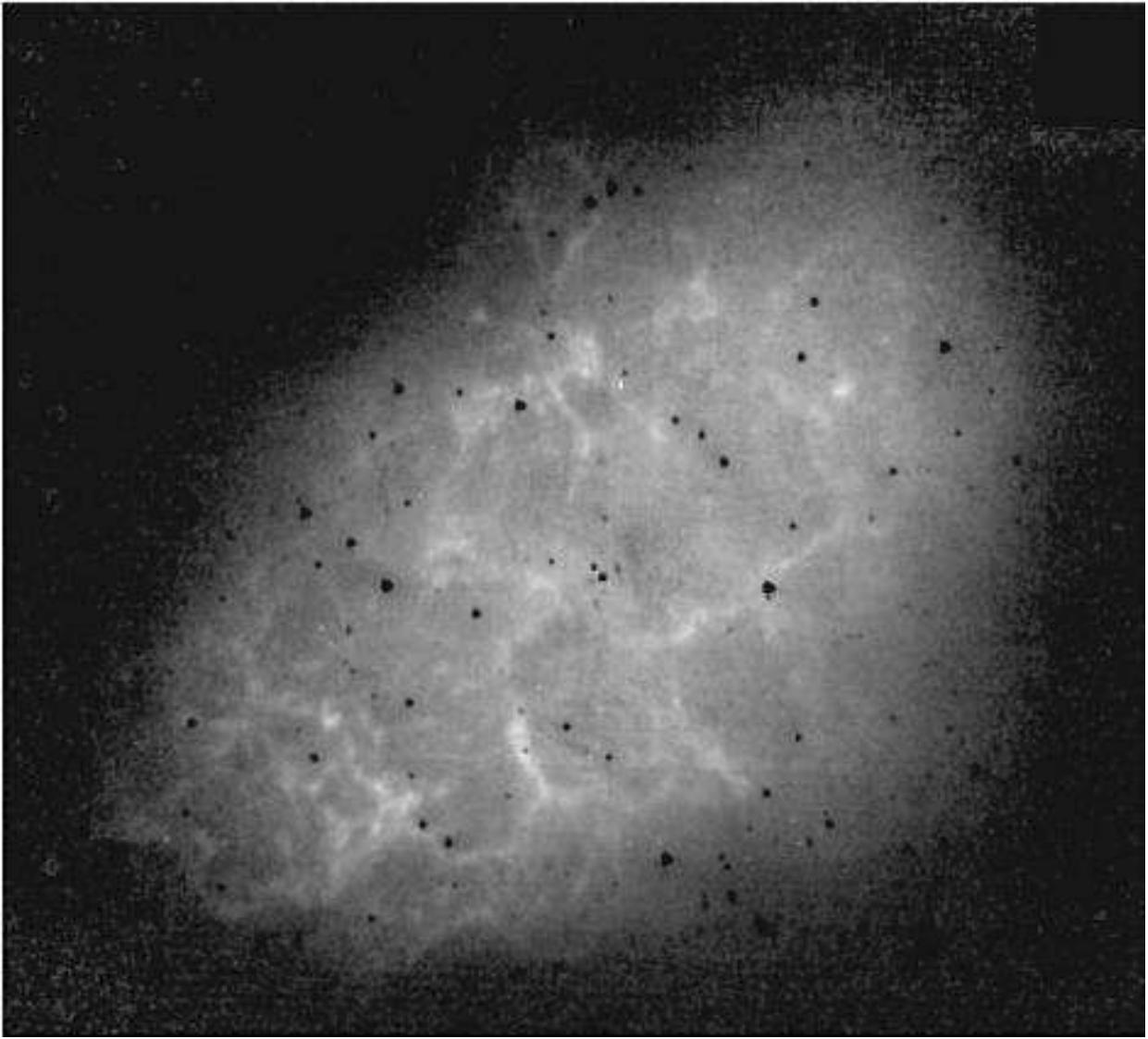} \caption{\label{residsynch} This
figure shows the difference between IRAC 3.6 $\micron$ and 5.8
$\micron$ images, where the 3.6 $\micron$ image has been scaled by a
factor of 1.3. Note that the jet and torus, which are prominent in
the original images are gone.}
\end{figure}

Neither the JP or KP models, nor the hybrid model developed above
can explain this shorter wavelength data. The basic problem is that
an exponential cutoff is needed, but it cannot follow a standard JP
shape, because any spectral steepening seen in the IR would predict
X-ray fluxes orders of magnitude lower than observed. So the
exponential cutoff must somehow be pushed out to shorter wavelengths
than expected. We empirically constructed such a model by modifying
the optical/IR ``hybrid'' to introduce a ``bump'' just below the
X-rays, followed by an exponential cutoff. This empirical  model is
labeled as hybrid 2 and shown both in color-color space (Figure
\ref{ccoptX}) and in normal log(brightness) {\em vs} log(frequency)
space in Figure \ref{crabspec2}. It starts at long wavelengths (low
energies) with the radio spectral index and fits the variations in
the IR-optical and optical-X-ray spectral indices.

Although this model is not unique, it has two characteristics that
must be present in any model that fits the data. First, the spectra
cut off very rapidly (possibly exponentially) in the X-ray regime,
indicative of strong radiative losses. Second, there needs to be a
``bump'' in the spectrum somewhere in the ultraviolet. The shaded
box in Figure \ref{crabspec2} shows the region of uncertainty from
our analysis; no direct measurements are available here. However,
without some fairly abrupt offset to the smooth spectral shape,
models including X-ray data would fail as badly as the JP and KP
models shown in Figure \ref{ccoptX}. Such an offset or bump in the
spectrum is expected due to the pileup of relativistic electrons
that result from energy losses at higher energies
\citep[e.g.][]{Rey03}, when the low frequency spectral index is
flatter than 0.5, as is true for the Crab. To our knowledge, there
has not been any previous observational evidence for this pileup.

\section{CONCLUSIONS \label{concl}}

We briefly summarize the main conclusions presented by our Spitzer
Space Telescope IR imaging and spectroscopic observations of the
Crab Nebula:

1.  A comparison of the morphology from the x-ray to the radio shows
that the synchrotron component and filaments dominate at different
wavelengths.

2.  We have derived a broadband shape to the synchrotron spectrum
which shows evidence for multiple components along the line of
sight, likely due to the torus and jet which remain flat to large
distances from the pulsar, superposed on the broad nebular emission
that steepens with increasing distance. At shorter wavelengths, the
derived spectral shape is consistent with the expected pileup of
relativistic electrons just below an exponential cutoff in X-rays.

3. We have measured the flux density of the nebula and the pulsar.
The smooth background of the nebula and the pulsar are dominated by
synchrotron emission and a large fraction of the emission from the
filaments in the images is due to forbidden line emission from Ar,
Ne, O, and Fe.

4. We find a paucity of dust. The small grain component seems to be
missing entirely and we see no evidence for silicate emission.  The
total emission at long wavelengths from large grains implies a total
dust mass in the nebula of less than 1\% of a solar mass.

5.  In the IRS spectra, we see Doppler shifted emission from both
the front and back sides of the expanding shell, and we measure a
radial expansion velocity of roughly 1264 km s$^{-1}$.

\acknowledgments

Support for this work was provided by NASA through contracts 1256406
and 1215746 issued by JPL/Caltech to the University of Minnesota. T.
Roellig acknowledges support from the NASA Office of Space Science.
We thank Matteo Murgia for providing the theoretical synchrotron
spectral distributions from his SYNMOD package and Steve Reynolds
and Bryan Gaensler for useful conversations about the status of
particle acceleration models. N.S. was supported by NASA through
grant HF-01166.01A from the Space Telescope Science Institute, which
is operated by AURA, Inc., under NASA contract NAS5-26555. This
publication makes use of data products from the Two Micron All Sky
Survey, which is a joint project of the University of Massachusetts and the
Infrared Processing and Analysis Center/California Institute of Technology,
funded by the National Aeronautics and Space Administration and the
National Science Foundation.

\bibliographystyle{plain}

\newpage
\begin{thebibliography}{}

\bibitem[Arendt et al.(1999)]{Are99} Arendt, R.~G., Dwek, E., \& Moseley, S.~H.\ 1999, \apj,
521, 234


\bibitem[Bandiera et al.(1998)]{Ban98} Bandiera, R., Amato,
E., \& Woltjer, L.\ 1998, Memorie della Societa Astronomica
Italiana, 69, 901

\bibitem[Bandiera et al.(2002)]{Ban02} Bandiera, R., Neri,
R., \& Cesaroni, R.\ 2002, \aap, 386, 1044

\bibitem[Bietenholz et al.(2001)]{Bie01} Bietenholz, M.~F.,
Frail, D.~A., \& Hester, J.~J.\ 2001, \apj, 560, 254

\bibitem[Bietenholz et al.(2004)]{Bie04} Bietenholz, M.~F.,
Hester, J.~J., Frail, D.~A., \& Bartel, N.\ 2004, \apj, 615, 794

\bibitem[Blair et al.(1997)]{Bla97} Blair, W.~P., Davidson,
K., Fesen, R.~A., Uomoto, A., MacAlpine, G.~M., \& Henry, R.~B.~C.\
1997, \apjs, 109, 473

\bibitem[Burbidge et al.(1957)]{Bur57} Burbidge, E.~M.,
Burbidge, G.~R., Fowler, W.~A., \& Hoyle, F.\ 1957, Reviews of
Modern Physics, 29, 547


\bibitem[Chevalier \& Gull(1975)]{Che75} Chevalier, R.~A., \&
Gull, T.~R.\ 1975, \apj, 200, 399

\bibitem[Clayton (1982)]{Cla82} Clayton, D.~D.\ 1982, \qjras,
23, 174



\bibitem[Davidson \& Fesen(1985)]{Dav85} Davidson, K., \&
Fesen, R.~A.\ 1985, \araa, 23, 119

\bibitem[Douvion et al.(2001)]{Dou01} Douvion, T., Lagage,
P.~O., Cesarsky, C.~J., \& Dwek, E.\ 2001, \aap, 373, 281

\bibitem[Dwek(1988)]{Dwe88} Dwek, E.\ 1988, \apj, 329, 814

\bibitem[Dwek(2004)]{Dwe04} Dwek, E.\ 2004, \apj, 607, 848


\bibitem[Eikenberry et al.(1997)]{Eik97} Eikenberry, S.~S.,
Fazio, G.~G., Ransom, S.~M., Middleditch, J., Kristian, J., \&
Pennypacker, C.~R.\ 1997, \apj, 477, 465

\bibitem[Fazio et al.(2004)]{Faz04} Fazio, G.~G., et al.\
2004, \apjs, 154, 10

\bibitem[Fesen \& Blair (1990)]{Fes90} Fesen, R., \& Blair,
W.~P.\ 1990, \apjl, 351, L45

\bibitem[Fesen \& Kirshner(1982)]{Fes82} Fesen, R.~A., \&
Kirshner, R.~P.\ 1982, \apj, 258, 1

\bibitem[Fesen et al.(1992)]{Fes92} Fesen, R.~A., Martin,
C.~L., \& Shull, J.~M.\ 1992, \apj, 399, 599


\bibitem[Gehrz(1988)]{Geh88} Gehrz, R.~D.\ 1988, \nat, 333,
705


\bibitem[Gehrz \& Ney(1987)]{Geh87}Gehrz, R. D., \&  Ney, E. P. 1987, Proc. Nat. Acad. Sci.
(USA), 84, 6961

\bibitem[Gehrz \& Ney(1990)]{Geh90}Gehrz, R. D., \&  Ney, E. P. 1990, Proc. Nat. Acad. Sci.
(USA), 97, 4354

\bibitem[Gehrz et al.(1998)]{Geh98} Gehrz, R.~D., Truran,
J.~W., Williams, R.~E., \& Starrfield, S.\ 1998, \pasp, 110, 3

\bibitem[Graham et al.(1990)]{Gra90} Graham, J.~R., Wright,
G.~S., \& Longmore, A.~J.\ 1990, \apj, 352, 172

\bibitem[Grasdalen(1979)]{Gra79} Grasdalen, G.~L.\ 1979,
\pasp, 91, 436

\bibitem[Green et al.(2004)]{Gre04} Green, D.~A., Tuffs,
R.~J., \& Popescu, C.~C.\ 2004, \mnras, 355, 1315

\bibitem[Gull \& Fesen(1982)]{Gul82} Gull, T.~R., \& Fesen,
R.~A.\ 1982, \apjl, 260, L75

\bibitem[Hartigan et al.(2004)]{Har04} Hartigan, P., Raymond,
J., \& Pierson, R.\ 2004, \apjl, 614, L69

\bibitem[Hester et al.(1990)]{Hes90} Hester, J.~J., Graham,
J.~R., Beichman, C.~A., \& Gautier, T.~N.\ 1990, \apj, 357, 539

\bibitem[Hester et al.(1995)]{Hes95} Hester, J.~J., et al.\
1995, \apj, 448, 240

\bibitem[Hester et al.(2002)]{Hes02} Hester, J.~J., et al.\
2002, \apjl, 577, L49

\bibitem[Higdon et al.(2004)]{Hig04} Higdon, S.~J.~U., et
al.\ 2004, \pasp, 116, 975

\bibitem[Houck et al.(2004)]{Hou04} Houck, J.~R., et al.\
2004, \apjs, 154, 18

\bibitem[Jaffe \& Perola(1973)]{Jaf73} Jaffe, W.~J., \&
Perola, G.~C.\ 1973, \aap, 26, 423

\bibitem[Kardashev(1962)]{Kar62} Kardashev, N.~S.\ 1962,
Soviet Astronomy, 6, 317

\bibitem[Katz-Stone et al.(1993)]{Kat93} Katz-Stone, D.~M.,
Rudnick, L., \& Anderson, M.~C.\ 1993, \apj, 407, 549

\bibitem[Kennel \& Coroniti(1984)]{Ken84} Kennel, C.~F., \&
Coroniti, F.~V.\ 1984, \apj, 283, 710




\bibitem[Lyubarsky(2002)]{Lyu02} Lyubarsky, Y.~E.\ 2002,
\mnras, 329, L34

\bibitem[MacAlpine et al.(1989)]{Mac89} MacAlpine, G.~M.,
McGaugh, S.~S., Mazzarella, J.~M., \& Uomoto, A.\ 1989, \apj, 342,
364

\bibitem[Marsden et al.(1984)]{Mar84} Marsden, P.~L.,
Gillett, F.~C., Jennings, R.~E., Emerson, J.~P., de Jong, T., \&
Olnon, F.~M.\ 1984, \apjl, 278, L29


\bibitem[Morse(2006)]{Mor06} Morse, J.A., Smith, N., Blair, W.P., Kirshner, R.P., Winkler,
P.F., \& Hughes, J.P.\ 2006, ApJ, in press.

\bibitem[Nodes et al.(2004)]{node04}
Nodes, C., Birk, G.~T., Gritschneder, M. \& Lesch, H. 2004, A\&A
423, 13

\bibitem[O'Connor et al.(2005)]{OCon05} O'Connor, P., Golden,
A., \& Shearer, A.\ 2005, \apj, 631, 471

\bibitem[Oliva et al.(1999)]{Oli99} Oliva, E., Moorwood,
A.~F.~M., Drapatz, S., Lutz, D., \& Sturm, E.\ 1999, \aap, 343, 943

\bibitem[Percival et al.(1993)]{Per93} Percival, J.~W., et
al.\ 1993, \apj, 407, 276

\bibitem[Reach et al.(2005)]{Rea05} Reach, W.~T., et al.\
2005, \pasp, 117, 978

\bibitem[Reynolds (2003)]{Rey03} Reynolds, S. astro-ph/0308483

\bibitem[Rieke et al.(2004)]{Rie04} Rieke, G.~H., et al.\
2004, \apjs, 154, 25

\bibitem[Rieke \& Lebofsky(1985)]{Rie85} Rieke, G.~H., \&
Lebofsky, M.~J.\ 1985, \apj, 288, 618

\bibitem[Sankrit et al.(1998)]{San98} Sankrit, R., et al.\
1998, \apj, 504, 344

\bibitem[Seward et al.(2006)]{Sew06} Seward, F. D., Gorenstein, P., \& Smith, R. K. 2006, ApJ,
636, 873.


\bibitem[Smith(2003)]{Smi03} Smith, N.\ 2003, \mnras, 346,
885

\bibitem[Sollerman(2003)]{Sol03} Sollerman, J.\ 2003, \aap,
406, 639

\bibitem[Sugerman et al.(2002)]{Sug02} Sugerman, B.~E.~K.,
Lawrence, S.~S., Crotts, A.~P.~S., Bouchet, P., \& Heathcote, S.~R.\
2002, \apj, 572, 209

\bibitem[Swinbank(1980)]{Swi80} Swinbank, E.\ 1980, \mnras,
193, 451

\bibitem[Uomoto \& MacAlpine(1987)]{Uom87} Uomoto, A., \&
MacAlpine, G.~M.\ 1987, \aj, 93, 1511

\bibitem[Veron-Cetty \& Woltjer(1993)]{Ver93} Veron-Cetty,
M.~P., \& Woltjer, L.\ 1993, \aap, 270, 370

\bibitem[Werner et al.(2004)]{Wer04} Werner, M.~W., et al.\
2004, \apjs, 154, 1

\bibitem[Willingale et al.(2001)]{willing} Willingale, R. et al. 2001, A\&A 365, L212


\end{thebibliography}

\begin{deluxetable}{rrccccccc}
\tablecolumns{9} \tablewidth{0pc} \tablecaption{IRAC and MIPS Data}
\tablehead{
 \multicolumn{2}{c}{} &  \colhead{} &  \colhead{} & \multicolumn{2}{c}{Nebula Flux Density} & \colhead{} & \multicolumn{2}{c}{Pulsar Flux Density} \\
 \cline{5-6} \cline{8-9} \\
\multicolumn{2}{c}{Wavelength}    &  \colhead{Exposure} & \colhead{Pixel} & \colhead{Raw } & \colhead{Extinction} & \colhead{} & \colhead{Raw}& \colhead{Extinction} \\
 \multicolumn{2}{c}{} &  \colhead{Time} & \colhead{Scale} & \colhead{} & \colhead{Corrected} & \colhead{} & \colhead{} & \colhead{Corrected} \\
 \multicolumn{2}{c}{($\micron$)} & \colhead{(s)}   & \colhead{(\arcsec/pix)} & \colhead{(Jy)} & \colhead{(Jy)} & \colhead{} & \colhead{(mJy)} & \colhead{(mJy)}
}
\startdata
IRAC & 3.6 & 12 & 0.86 & 11.65 $\pm$ 0.03 & 12.63 $\pm$ 0.22 & & 2.45 $\pm$ 0.16 & 2.66 $\pm$ 0.19  \\
& 4.5 & 12 & 0.86 & 13.81 $\pm$ 0.02 & 14.35 $\pm$ 0.26 & & 2.34 $\pm$ 0.17 & 2.42 $\pm$ 0.18 \\
& 5.8 & 12 & 0.86 & 16.31 $\pm$ 0.06 & 16.77 $\pm$ 0.13 & & 2.12 $\pm$ 0.19 & 2.18 $\pm$ 0.19 \\
& 8.0 & 12 & 0.86 & 17.74 $\pm$ 0.04 & 18.33 $\pm$ 0.13 & & 1.85 $\pm$ 0.18 & 1.91 $\pm$ 0.19 \\
MIPS & 24.0 & 3 & 2.5 &  59.8 $\pm$ 0.4  & \nodata & & \nodata & \nodata \\
& 70.0 & 10 & 4.0 & 157 $\pm$ 2 & \nodata & & \nodata & \nodata \\
\enddata
\label{irac_mips}
\end{deluxetable}

\begin{deluxetable}{crcc}
 \tablecolumns{4} \tablewidth{0pc} \tablecaption{IRS
Observations} \ \tablehead{
\colhead{IRS Module Name} &  \colhead{Spectral} &  \colhead{Spectral} & \colhead{Integration Time at} \\
\colhead{} &  \colhead{Range} &  \colhead{(Resolution} & \colhead{IRS-Tgt-Cntr} \\
\colhead{} &  \colhead{($\micron$)} &  \colhead{($\lambda$/${\Delta}{\lambda}$)} & \colhead{(sec)}
}
\startdata
Short-Low (SL) & 5.2 - 14.3 & 64 - 128  &  928 \\
Long-Low (LL) & 14.1 - 38.0 & 64 - 128  &  488 \\
Short-High (SH) & 9.9 - 19.6 & $\sim$ 600 & 1768 \\
Long-High (LH) & 18.9 - 37.2 & $\sim$ 600 &  1936 \\
\enddata
\label{irs}
\end{deluxetable}

\begin{deluxetable}{ccccc}
\tablecolumns{5} \tablewidth{0pc} \tablecaption{IRS Short-High Data} \
\tablehead{
\colhead{Line} &  \colhead{Wavelength} &  \colhead{Measured} & \colhead{Line Fit Height} & \colhead{Line Flux} \\
\colhead{} &  \colhead{($\micron$)} &  \colhead{Wavelength ($\micron$)} & \colhead{(Jy)} & \colhead{(W/cm$^2$/$\micron$)}
}
\startdata
[NeII] & 12.8135 & 12.763 $\pm$ 0.001 & 0.6850 $\pm$ 0.0044 & 4.386 $\pm$ 0.043 e-20 \\
& & 12.862 $\pm$ 0.001 & 0.1252 $\pm$ 0.0038 & 8.930  $\pm$ 0.048 e-21 \\

[NeV] &  14.3217 & 14.261  $\pm$ 0.007 & 0.0123 $\pm$ 0.0035 & 9.325 $\pm$ 3.311 e-22 \\
& & 14.404  $\pm$ 0.008 & 0.0099 $\pm$ 0.0035 & 7.851 $\pm$ 4.603 e-22 \\

[NeIII] & 15.5551 & 15.493  $\pm$ 0.001 & 0.9434 $\pm$ 0.0556 & 4.692 $\pm$ 0.039 e-20 \\
& & 15.621  $\pm$ 0.001 & 0.2982 $\pm$ 0.0038 & 1.900 $\pm$ 0.039 e-20 \\

[SIII] & 18.7130 & 18.648  $\pm$ 0.001 & 0.0475 $\pm$ 0.0025 & 2.533 $\pm$ 0.201 e-21 \\
& & 18.777  $\pm$ 0.002 & 0.0483 $\pm$ 0.0025 & 2.452 $\pm$ 0.197 e-21 \\
\label{sh}
\enddata
\end{deluxetable}

\begin{deluxetable}{cccccc}
  \tablecolumns{6} \tablewidth{0pc} \tablecaption{IRS Long-High Data} \ \tablehead{
\colhead{Line} &  \colhead{Wavelength} &  \colhead{Measured} &
\colhead{Line Fit Height} & \colhead{Line Flux}&\colhead{Scaled Line
Flux$^{a}$} \\
\colhead{} &  \colhead{} &  \colhead{Wavelength} & \colhead{} & \colhead{} &
\colhead{} \\
\colhead{} &  \colhead{($\micron$)} &  \colhead{($\micron$)} & \colhead{(Jy)} & \colhead{(W/cm$^2$/$\micron$)} &
\colhead{(W/cm$^2$/$\micron$)}
} \startdata
  [FeIII]& 22.9250 & 22.801 $\pm$ 0.021 & 0.030 $\pm$ 0.009  & 1.28
$\pm$ 0.6 e-21
  &4.2 $\pm$ 1.9 e-22 \\
  &  & 23.004 $\pm$ 0.024 & 0.028 $\pm$ 0.007 & 1.73 $\pm$ 0.7 e-21
  & 5.7 $\pm$ 2.2 e-22 \\

[NeV] & 24.3175 & 24.210 $\pm$ 0.005 & 0.140 $\pm$ 0.007 & 9.33 $\pm$ 0.7 e-21
& 3.1 $\pm$ 0.2 e-21 \\
& & 24.420 $\pm$ 0.006 & 0.114 $\pm$ 0.007 & 6.26 $\pm$ 0.6 e-21
&  2.1 $\pm$ 0.2 e-21 \\

[OIV]+ & 25.8903& 25.772 $\pm$ 0.006 &  1.372 $\pm$ 0.007& 9.13 $\pm$ 0.07 e-20
& 3.08 $\pm$ 0.02 e-20 \\
weak [FeII]& & 25.999 $\pm$ 0.006&  1.342 $\pm$ 0.007& 8.28 $\pm$ 0.06 e-20
&  2.73 $\pm$ 0.02 e-20\\

[SIII] & 33.4810 & 33.326 $\pm$ 0.002 & 0.376 $\pm$ 0.025 & 2.07 $\pm$ 0.2 e-20
& 6.8 $\pm$ 0.7 e-21 \\
& & 33.547 $\pm$ 0.001 & 0.479 $\pm$ 0.028 & 2.40 $\pm$ 0.2 e-20
& 7.9 $\pm$ 0.8 e-21 \\

[SiII]&34.8152 & 34.675 $\pm$ 0.001 & 0.780 $\pm$ 0.032 & 3.24 $\pm$ 0.2 e-20
& 1.07 $\pm$ 0.07 e-20\\
& & 34.934 $\pm$ 0.002 & 0.383 $\pm$ 0.033 & 1.30 $\pm$ 0.2 e-20
& 4.3 $\pm$ 0.6 e-21\\

[FeII]& 35.3487& 35.194 $\pm$ 0.007 & Same as below & \nodata & \nodata\\
& & 35.455 $\pm$ 0.009 & 0.093 $\pm$ 0.015 & 3.71 $\pm$ 2.1 e-21
& 1.2 $\pm$ 0.7 e-21\\

[NeIII]& 36.0135& 35.852 $\pm$ 0.001 & 0.865 $\pm$ 0.034 &  2.50 $\pm$ 0.2 e-20
&8.2 $\pm$ 0.6 e-21\\
& & 36.138 $\pm$ 0.001 & 0.513 $\pm$ 0.031 & 1.74 $\pm$ 0.2 e-20
& 5.8 $\pm$ 0.6 e-21\\
\label{lh}
\enddata
\tablenotetext{a}{The uncertainties in the displayed scaled line
fluxes are statistical only. There are additional large systematic
uncertainties in the scale factor -- see the text for details.}
\end{deluxetable}

\clearpage

\clearpage

\clearpage

\clearpage

\clearpage

\clearpage

\clearpage

\clearpage

\clearpage

\clearpage

\clearpage

\clearpage

\clearpage

\clearpage

\clearpage

\end{document}